\documentclass[preprint,aps,nofootinbib]{revtex4}
\usepackage{dcolumn}% Align table columns on decimal point
\usepackage{bm}% bold math
\usepackage{tabularx}
\usepackage{amsmath}
\usepackage{amscd}
\usepackage{delarray}
\usepackage{amssymb}
\usepackage{graphicx}

\newcommand{\rr}{\sqrt{2}}

\newcommand{\psla}{p\kern-.45em/}
\newcommand{\ebsla}{\bar{\epsilon}\kern-.45em/}
\newcommand{\qsla}{q\kern-.45em/}
\newcommand{\pbsla}{\bar{p}\kern-.45em/}
\newcommand{\qbsla}{\bar{q}\kern-.45em/}

\newcommand{\slush}{\!\!\!/}

\def\ti    {\tilde}

\def\none  {\ti \chi^0_1}

\def\h     {\frac{1}{2}}

\begin{document}

\preprint{KEK--TH--1233}
\preprint{IPMU08--0010}
\preprint{YITP--08--10}

\title{Study of the top reconstruction 
in top-partner events at the LHC}

\author{Mihoko M.Nojiri}
 \email{nojiri@post.kek.jp}
\author{Michihisa Takeuchi}%
 \email{tmichihi@post.kek.jp} 
\affiliation{$^{\ast\dagger}\!\!\!$
Theory Group, KEK,
and $^{\ast}\!\!\!$ the Graduate University for Advanced Studies (SOKENDAI)\\
1-1 Oho, Tsukuba, 305-0801, Japan\\
$^{\ast}\!\!\!$ IPMU, Tokyo University, Kashiwa, Chiba, 277-8568, Japan\\
$^\dagger\!\!\!$ Yukawa Institute for Theoretical Physics, Kyoto University,\\
 Kyoto 606-8502, Japan
}

\date{\today}

\begin{abstract}
In the Littlest Higgs model with T-parity (LHT),
top-partners ($T_-$) are produced in pairs
at the Large Hadron Collider (LHC).  
Each top-partner decays into a top quark ($t$) and the lightest 
$T$-odd gauge partner $A_H$.
We demonstrate reconstruction of the  $t\bar{t}$ system decaying 
hadronically, and measurement of the top-partner mass 
from the  $m_{T2}$ distribution. 
A top quark from a $T_{-}$  decay is polarized, and we discuss the effect of this polarization
on the decay distributions. 
Because the events consist of  highly collinear jets  which 
occasionally overlap each other,  
we  compare distributions using different jet reconstruction 
algorithms (Snowmass cone, kt, Cambridge, SISCone). 
We find clustering algorithms are advantageous  
for studying top polarization effects.
\end{abstract}

\maketitle

%\pacs{Valid PACS appear here}

%\newpage

\section{Introduction}
\label{sec:introduction}
In spite of the success of the Standard Model (SM)
in explaining the particle interactions, 
there are yet two unsolved questions in the SM. 
One  is the fine tuning problem and the other is 
the existence of the dark matter (DM) in our universe.

The fine tuning problem is the question 
why the Higgs is likely to be so light 
as expected from the LEP data ($m_{h}<198{\rm~GeV}$ at the 95\% confidence level
 \cite{Barate:2003sz}),
while naturally the Higgs mass is the order of the cut off scale of the theory
due to radiative corrections.
Some mechanism should protect the Higgs mass, and interactions involving 
the Higgs sector should be extended from the SM one.

The existence of the DM is now established  by the 
cosmological  observations such as WMAP, SDSS and SN-Ia \cite{wmap,wmap2,York:2000gk,Perlmutter:1998np,Riess:1998cb}.
The SM should be extended to include the DM, which should be
a neutral stable particle. 
Moreover, provided that the DM is a thermal relic whose
strength of the coupling to the SM particles of the order of the weak interaction, 
the mass  can be  a few 100~GeV from a rough dimensional analysis. 

To solve the fine tuning problem,
many models have been proposed.
However, a so-called \lq\lq little hierarchy problem"
\cite{Barbieri:2000gf,Barbieri:1999tm} 
arises in these models, 
once constraints from precision measurements are imposed. 
New operators arising from these models 
naively have the cut off scale $\Lambda > 10$ TeV
to be consistent to  the experimental data \cite{Csaki:2002qg}.

Some successful models solving the little hierarchy problem
have parity structures,
for example the MSSM with R-parity
\cite{Nilles:1983ge,Haber:1984rc,Martin:1997ns},
the little Higgs model with T-parity (LHT)
\cite{ArkaniHamed:1998rs,Antoniadis:1998ig,Cheng:2003ju,Cheng:2004yc,Low:2004xc},
or the universal extra dimensional model (UED) with KK-parity 
\cite{UED1,UED2}.
Such a model which has a parity structure predicts
the stable lightest parity odd particle
as a candidate of the DM.
The cut off scale of new operators in the model can be as low as  a few TeV \cite{Hubisz:2004ft}, 
and the mass of the lightest parity odd particle can be on the order of a few 100~GeV.

The LHC is starting soon,
and it is likely to discover new particles with masses up to a few TeV
\cite{atlas,cms}.
The models mentioned above are studied intensively by many authors.
All of these  models have  partners of  the SM colored particles 
which decay into the stable lightest parity odd particle 
through the parity conserving interactions.

At the  LHC, these partners are produced in pairs.
The signal is multiple high $p_T$ jets and high $p_T$ leptons,
each accompanied by missing transverse momentum $E\slush_T$.
Among the signals studied so far, 
the signals with high $p_T$ leptons are very promising \cite{Baer:1991xs,Baer:1995va},
because the SM backgrounds are smaller. 
However, branching ratios of new particles into leptons strongly depend on 
model parameters.
In addition,
Jets + leptons signals are often accompanied by undetectable  neutrinos, 
which are also the source of missing momentum.
They sometimes reduce the significance of the kinematical endpoints 
of the signal distributions for mass reconstructions. 
And even in the case that the lepton branching ratios are small, 
events with multiple jets and  no lepton  are enormously produced.

In this paper, we focus on the top partner signal in the LHT. 
The top partner pair productions occur
with sizable cross section \cite{Belyaev:2006jh} 
compared with stop pair productions in the MSSM 
because the top partner is a fermion.
The event has simple kinematics, $t\bar{t}+ E\slush_T$ where top quarks 
are highly boosted\footnote{Similar highly boosted $t\bar{t}$ signals have been discussed in the RS1 model~\cite{Lillie:2007yh}.
However, it is different from the signal considered in this paper,
because it is not accompanied by $E\slush_T$.}.
Decay products from a boosted top are collimated 
and they are easily identified as a jet system with mass 
$\sim m_t$ with high probability using hemisphere analysis~\cite{hemisphere,Matsumoto:2006ws}.
On the other hand, by imposing lepton veto and top tagging, 
background events from $t\bar{t}+$jets can be reduced significantly. 

The process above was partly studied in Ref.~\cite{Matsumoto:2006ws}. 
In the paper, {\tt COMPHEP} \cite{Pukhov:1999gg} and {\tt HERWIG6.5} \cite{Corcella:2002jc} 
are used for the event generation and
{\tt AcerDET1.0} \cite{Richter-Was:2002ch}  is used for the detector simulation and jet reconstruction.
{\tt AcerDET1.0} implements the Snowmass cone algorithm.
However,
the algorithm is not optimized in resolving the overlapping jets
arising from boosted top quark correctly.

Recently,  {\tt FastJet} \cite{Cacciari:2006sm} 
was released.
Infrared stable jet reconstruction algorithms (kt \cite{Catani:1991hj,Catani:1993hr,Ellis:1993tq,Cacciari:2005hq}, Cambridge \cite{Dokshitzer:1997in,Bentvelsen:1998ug}, 
SISCone \cite{Salam:2007xv}) are implemented in the code, which is significantly faster than previous codes. 
We study improvements with these advanced jet 
reconstruction algorithms. 
In this study, we interface {\tt AcerDET} calorimeter 
information to  {\tt FastJet} and reanalyze the same process as 
in Ref.~\cite{Matsumoto:2006ws} to compare the results.
We find the kt and Cambridge algorithm have advantage 
to resolve overlapping jets.
We also generate the signal events 
with underlying events
using {\tt HERWIG6.5} + {\tt Jimmy} \cite{Butterworth:1996zw},
and find the jet resolution with the kt algorithm 
is  significantly affected
by underlying events.
This motivates us to show 
results with the Cambridge algorithm mainly in this paper. 

We study the potential to measure the mass of the top partner
 using the reconstructed top candidates,
although only the discovery potential is discussed in Ref.~\cite{Matsumoto:2006ws}.
One of the important variable is a $m_{T2}$ \cite{Barr:2003rg}. 
The $m_{T2}$ is a function of two visible momenta, a missing 
transverse momentum and a test mass.
In the case that the mass of the lightest $T$-odd particle ($m_{\rm LTP}$) is known, 
the endpoint of the $m_{T2}$ distribution is equal 
to the top partner mass at $M_{\rm test}=m_{\rm LTP}$.
Therefore we can measure the top partner mass $m_{T_-}$ using this distribution.
We also generate
the Standard Model backgrounds using {\tt ALPGEN} \cite{Mangano:2002ea}+ {\tt HERWIG} 
and conclude that they do not affect the endpoint of the $m_{T2}$ distribution.

We also discuss top polarization effects.
A typical LHT model predicts a top partner which decays dominantly 
into a right-handed top quark $t_R$ and a heavy photon $A_H$.
The polarization of tops can be measured through 
investigating decay distributions of tops.
And we show that there are distinguishable difference between 
completely polarized case and non-polarized case
in jet level analysis.

This paper is organized  as follows. 
In Sec.~\ref{lht}, 
we explain our simulation setup for studying a top partner at the LHC.
And we show how to reconstruct momenta of top quarks arising from top partner decays
and how to measure the top partner mass $m_{T_-}$ using a reconstructed $m_{T2}$ distribution.
In Sec.~\ref{toppartner_label2c}, 
we discuss differences among 
the jet reconstruction algorithms. 
In Sec.~\ref{toppolarization}, 
we study top polarization effects.
Sec.~\ref{conclusion} is devoted to the discussions and conclusions.

\section{Top partner reconstruction at the  LHC}
\label{lht}
\subsection{Event generation}
In the following, we assume the top partner is the lightest in the fermion partners 
and decays exclusively to the lightest $T$-odd particle $A_H$ and a top. 
The top partner may be produced in pairs at the LHC  and decays   as,
\begin{eqnarray}
pp \rightarrow T_-\overline{T}_- \rightarrow t\bar{t}A_H A_H 
\rightarrow bW^+\bar{b}W^- A_H A_H 
\rightarrow 6j + E{\!\!\!/}_T.
\end{eqnarray}
This process is similar to scalar top ($\tilde t$) pair production
process in the MSSM.
The production cross section of top partner
is larger than that of  scalar top in the case that the masses are the same, because
top partner is a fermion. 
At the LHC, $T_-\overline{T}_-$ production cross section is 0.171 pb
for  $m_{T_-}=800$~GeV.
In order to identify this process, it is important to tag  top quarks, and measure
missing transverse momentum $E{\!\!\!/}_T$ arising from escaping $A_{H}$'s.

In the previous study \cite{Matsumoto:2006ws},
the events $pp\rightarrow t\bar{t} A_H A_H$ 
are generated by  {\tt COMPHEP} \cite{Pukhov:1999gg},
and top quark momenta are interfaced to {\tt HERWIG6.5} \cite{Corcella:2002jc}. 
In this paper, we also use  {\tt COMPHEP}  and {\tt HERWIG6.5} for signal events generation. 
Our study is based on 8,550 signal events 
corresponding to $\int dt {\cal L}=50$fb$^{-1}$. 

The Standard Model background to the signal comes 
from the production of QCD, $t\bar{t}+ n$ jets, $W+n$ jets, and $Z+n$ jets events. 
 In ATLAS study, it was shown that these four  processes  contribute to the background 
 of the 0 lepton $+ E\slush_T +$ jets channel for SUSY search 
  with approximately the same order of  magnitude \cite{SUSY08deJongtalk}. 
Among those,  
the  QCD background arises due to the detector smearing and inefficiency, 
 and we do not attempt to simulate it in this paper.
Even if QCD background is taken into account,
it will not affect the results significantly because we require top mass cuts for the event selection.
We will discuss this point later.
The other processes contributes to the background due to hard $\nu$ 
produced from $Z$ and $W$ decay. They are generated by
{\tt ALPGEN}+{\tt HERWIG} in the paper.
To reduce the computational time, 
 we generate $Z(\rightarrow \nu\bar{\nu}) + n$ jets ($n\le 4$) events 
 corresponding to  5 fb$^{-1}$
 with $E\slush_T> 150$ GeV,  
 $W(\rightarrow l\nu)+n$ jets ($n\le 4$) corresponding to 10fb$^{-1}$  
 with $\sum_{\rm parton} E_T> 400$ GeV  , 
 and $t\bar{t} +n$ jets ($n\le 2$) corresponding to 
 12 fb$^{-1}$   with $\sum_{\rm parton} E_T> 500$ GeV   respectively.  
  Parton shower and matrix element matching are performed 
 using MLM scheme  provided by {\tt ALPGEN}.  We require 
$\eta_{\rm max}=5$, $p_{T{\rm min}}> 30$GeV and $R_{jj}>0.4$ 
for parton level  event  generation before the matching.   
 The $t\bar{t}Z$ followed by $Z\rightarrow \nu\bar{\nu}$ events become irreducible backgrounds, and we have generated the events for 
 roughly $50$fb$^{-1}$.   We do not apply K-factor both for signal and background.

We  use {\tt AcerDET1.0} for detector simulation,
 particle identification and  jet reconstruction as in \cite{Matsumoto:2006ws}. 
In addition, we interfaced  calorimeter information of {\tt AcerDET1.0} to
{\tt FastJet2.2beta} \cite{Cacciari:2006sm} so that we can compare different jet 
reconstruction algorithms simultaneously, and model the detector granularity.
Here the calorimeter information is the energy deposit  $E^{\rm sum}_{i}$
to each cell $i$ centered at ($\phi_i$, $\eta_i$) with the size  
 $\Delta\phi=0.1$ and $\Delta\eta=0.1$ (0.2 in the forward  directions) 
 without smearing.  They are interfaced as 
massless particles with momenta $p_i=(E_i^{\rm sum}, \eta_i, \phi_i$) to {\tt FastJet} 
\footnote{Effects of  shower propagation to nearby cells are not taken into account.}.
In this paper,  we study jet distributions in the infrared stable algorithms,
(kt, Cambridge and  SISCone), together in those for the  Snowmass cone algorithm provided by {\tt AcerDET}.
To compare the four jet algorithms under the same conditions in Sec.~\ref{toppartner_label2},
we switch off  the jet energy smearing.
For the background, the smearing on $E\slush_T$ might
have the same impact on the estimation of the number
of events after the cut, therefore we use the $E\slush_T$
smeared by the  {\tt AcerDET}. 
Effects of Jet energy smearing are discussed in the Appendix \ref{smearing}.

\subsection{Event selection and Top reconstruction}
\label{toppartner_label1}

\begin{table}[h]
\footnotesize
\begin{tabular}{|c||r|r||r|r|r|r||r|r|}
\hline
   & $ M_{\rm eff},E\slush_T$ ,  $n_{\rm lep}$ &  $p_{H_i}>200$ & $m_{P_{H1}}\!\!\sim\!\!  m_t$
& $ m_{P_{H2}} \!\!\sim\!\! m_t$&  $\ $ both$\ $ & $\ $or$\ \ $& $m_{T2}>350$& $m_{T2}>500$ \cr
\hline
\hline
$T_- \overline{T}_-$(signal)
&$     2,764$&$     1,675$&$      404$&$      396$&$      130$&$      398$&$      372$&$      199$\cr
\hline
$t\bar{t}+$jets
&$    34,906$&$    12,296$&$     2,114$&$     1,288$&$      241$&$     1,230$&$      192$&$        0$\cr
$t\bar{t}Z(\rightarrow \nu\bar{\nu})$ 
&$      337$&$       95$&$       16$&$       24$&$        5$&$       19$&$        3$&$        3$\cr
$Z(\rightarrow \nu \bar{\nu}) +$jets 
&$    26,290$&$     8,676$&$      520$&$      890$&$       50$&$      420$&$      280$&$       10$\cr
$W(\rightarrow \l \nu) +$jets 
&$    24,045$&$     7,780$&$      465$&$      700$&$       55$&$      285$&$      140$&$       10$\cr
\hline
\end{tabular}
\caption{Summary of the number of events after the cuts
 for  $\int\!\! dt {\cal L} = 50\,$ fb$^{-1}$. 
Numbers of BG events are properly scaled to 50 fb$^{-1}$.
See the text for the detail. 
}
\label{toppartner_label3}
\end{table}

We describe our cuts to select $T_-\overline{T}_-$ events. 
The summary of the numbers of the events after the cuts is shown 
in Table \ref{toppartner_label3}.
First, we impose our  standard cuts  for jet $p_T$,  $E\!\!\!/_T$ 
and veto high $p_T$ isolated leptons,
\begin{eqnarray}
E\!\!\!/_T \ge 200{\rm~GeV\ and}\ E\!\!\!/_T \ge 0.2M_{\rm eff},\ \ 
n_{50} \ge 4\ {\rm and}\  n_{100} \ge 1,\ \ 
n_{\rm lep}=0.
\label{basiccut}
\end{eqnarray}
Here,
\begin{eqnarray}
M_{\rm eff}=\sum_{p_T>50{\rm GeV} \atop |\eta|<3}p_T^{\rm jet}
+\sum_{p_T>10{\rm GeV} \atop |\eta|<2.5}p_T^{\rm lepton}
+\sum_{p_T>10{\rm GeV} \atop |\eta|<2.5}p_T^{\rm photon}
+E\slush_T,
\end{eqnarray}
$n_{50}\ (n_{100})$ is a number of jets whose $p_T$ is larger than 50 (100) GeV.
$n_{\rm lep}$ is a number of isolated leptons ($e,\mu$) with $p_T \ge 5 $ GeV and $|\eta|<2.5$.
Missing transverse energy is calculated using the energy deposit
to the calorimeter and isolated leptons.
It is calculated with smearing for the Standard Model background calculation.

The lepton cut reduces $t\bar{t} +$ jets  and $W+$ jets background,
in which large $E\!\!\!/_T$ is
dominantly caused by neutrinos from  leptonic $W$ decay.
$W$ background still remains because $W$ can decay into $\tau$.

We applied a hemisphere analysis to find top candidates \cite{hemisphere}. 
Each of high $p_T$ jets ($p_T > 30$~GeV and $\vert \eta \vert<3$ )
 is assigned to one of the two hemispheres which  are defined as follows;
\begin{eqnarray}
&&\forall i \in H_1, j \in H_2 \ \ \ \ \ \  \ \ d(p_{H_1},p_{i})\le d(p_{H_2},p_{i}) {\rm \ and\  }
d(P_{H_2},p_{j})\le d(P_{H_1},p_{j}),
\label{distance}
\end{eqnarray}
where
\begin{eqnarray}
 P_{H_i}&\equiv&\sum_{k\in H_i} p_k,
\label{hemidef}
\end{eqnarray}
\begin{eqnarray}
 d(p_{i},p_j)\equiv
\frac{(E_{i}-|\mathbf{p}_{i}|\cos\theta_{ij})E_{i}}{(E_{i} + E_{j})^2},
\label{hemidistancedef}
\end{eqnarray}
\begin{eqnarray}
\cos\theta_{ij}
\equiv \frac{\mathbf{p}_i \cdot \mathbf{p}_j}{|\mathbf{p}_i||\mathbf{p}_j|}.
\ \ \ \ \ 
(\theta_{ij}
\ {\rm is\  the\ angle\ between\ }
\mathbf{p}_i\ {\rm and}\ \mathbf{p}_j).
\end{eqnarray}

To find hemispheres, we first take the highest $p_T$ jet momentum $p_1$ as $P_{H_1}$
and take the jet momentum $p_k$ 
which maximizes $\Delta R(p_1,p_k) \cdot p_{kT}$ among all $k$  as  $p_{H_2}$.
We group jets into hemisphere $H_i$ $(i=1,2)$ according to the eq.(\ref{distance}).
New $P_{H}$'s are then calculated from Eq.(\ref{hemidef}), and
this process is repeated until the assignment converges. 
In this analysis, collinear objects  tend to be assigned into the same hemisphere. 
Top quarks from $T_-$ decays 
are highly boosted, then the decay products from the two top quarks  are correctly
grouped into different hemispheres with  high probability.
In this situation,
the dependence on the definition of the distance eq.(\ref{hemidistancedef})
is weak.
Change the definition of the distance as $d(p_i,p_j)\equiv\theta_{ij}$
 causes negligible differences of the acceptance and our analyses in this paper.

To assure the correct top reconstructions, 
 we require both hemispheres' transverse momenta are larger than a threshold,
\begin{eqnarray}
 P_{T,H_1},P_{T,H_2} > 200 {\rm ~GeV}.
\label{ptcut}
\end{eqnarray}

\begin{figure}[h!]
\includegraphics[scale=0.4]{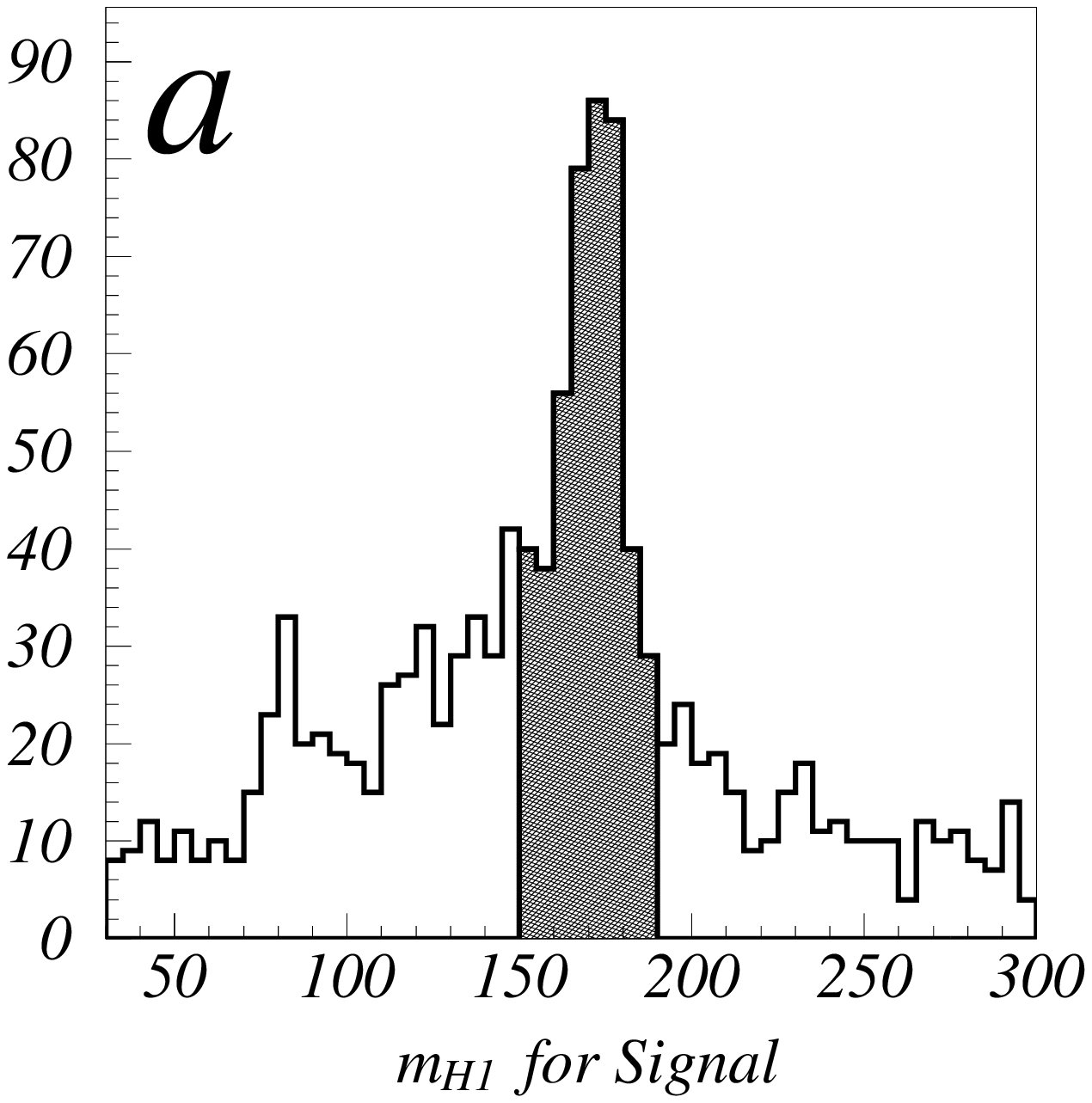}
\includegraphics[scale=0.4]{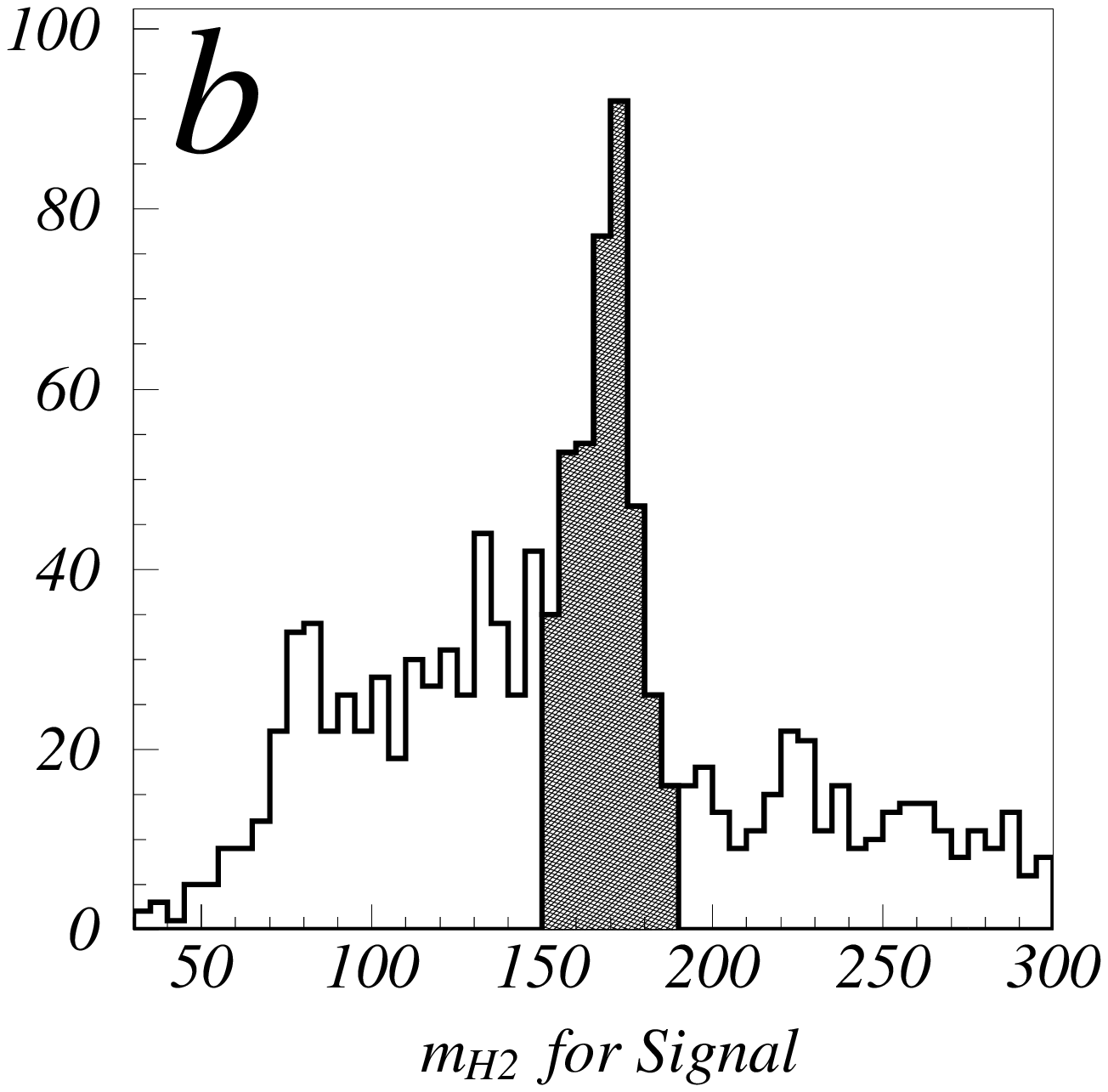}
\\
\includegraphics[scale=0.4]{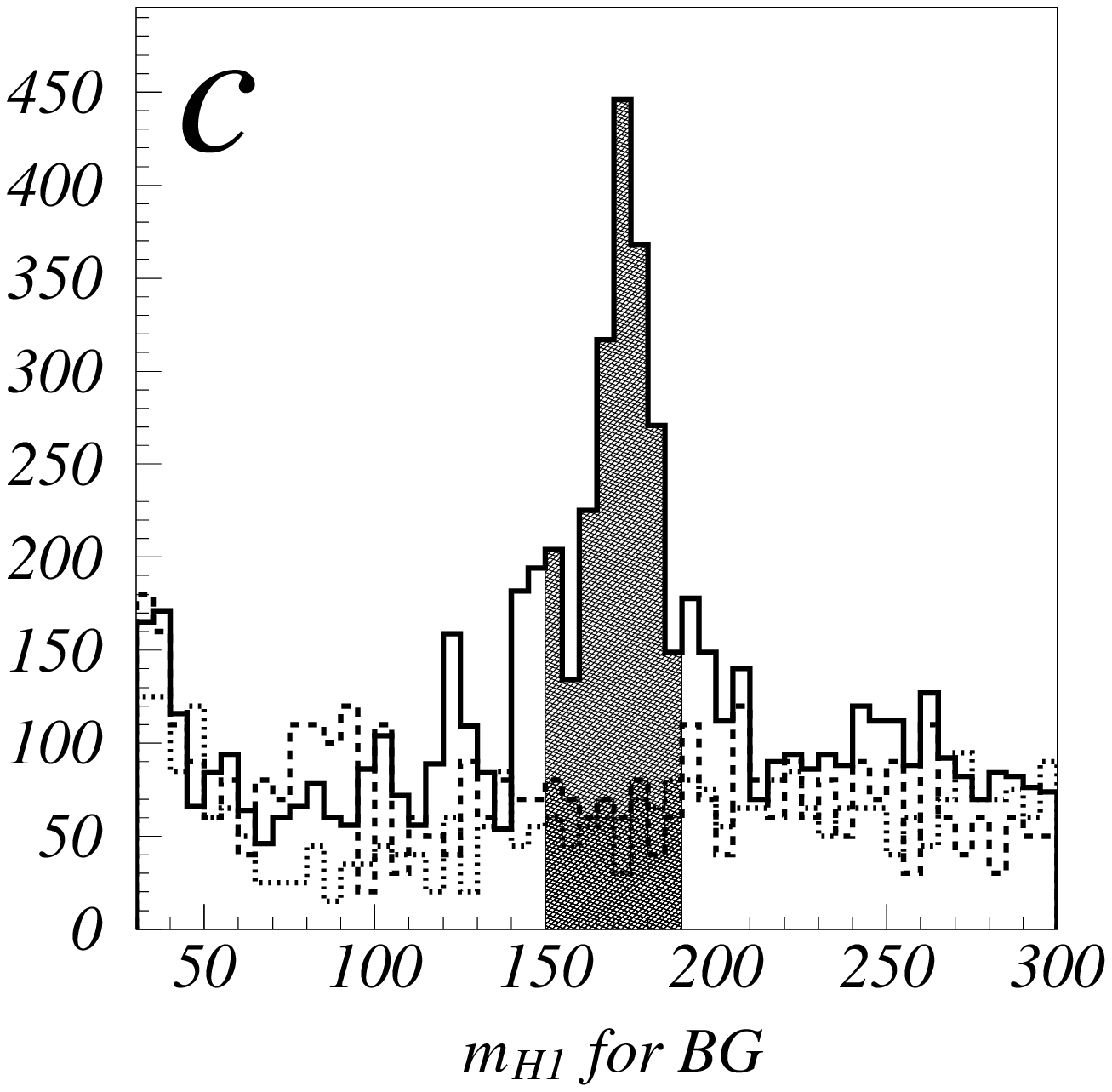}
\includegraphics[scale=0.4]{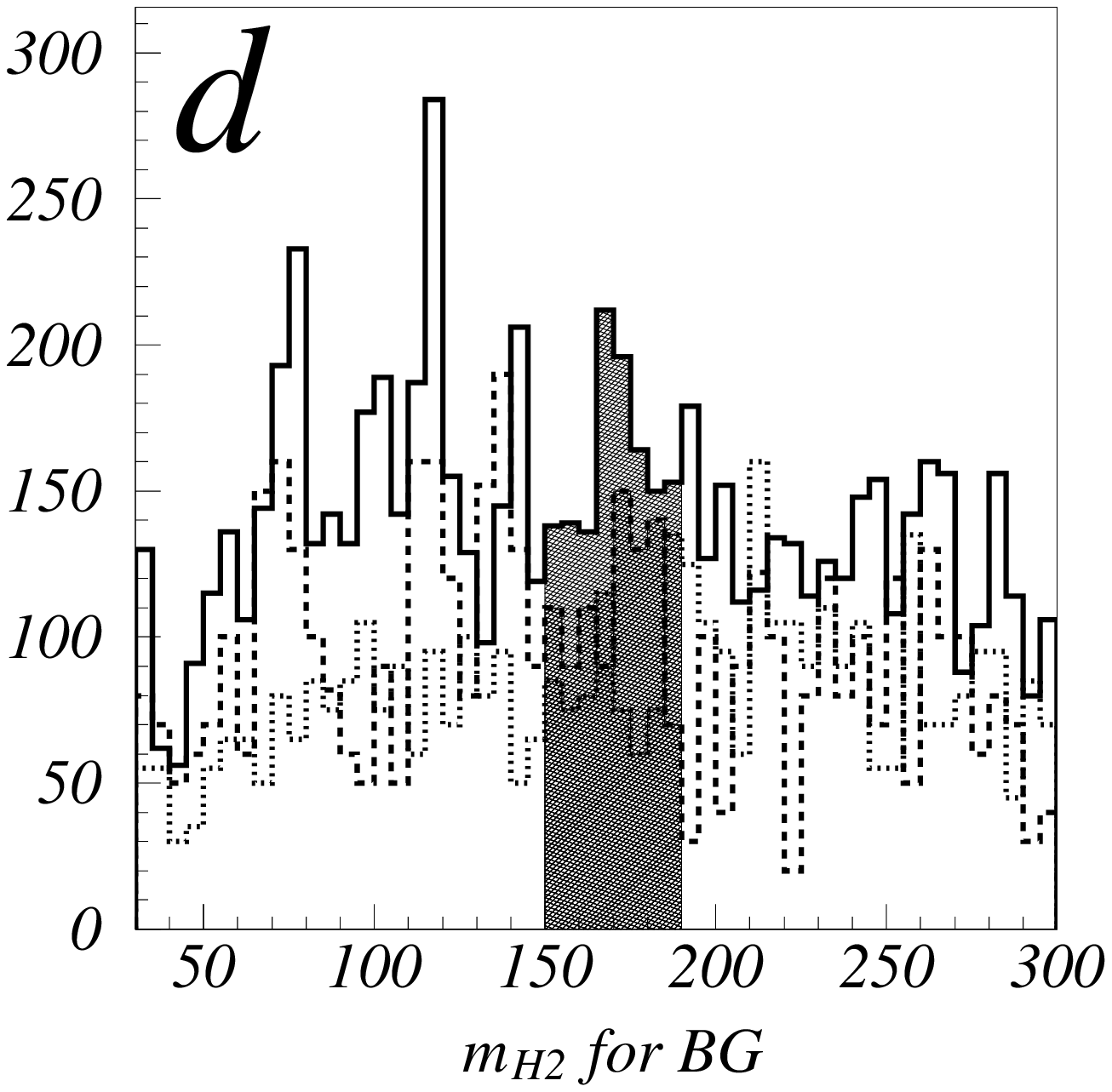}
\caption{The distributions of the invariant masses of 
a) $H_1$  and b) $H_2$ for the $T_-\overline{T}_-$ events for 50 fb$^{-1}$,
c) $H_1$  and d) $H_2$ for the events from  $t\bar{t}+$jets (solid) , $W+$jets (dotted)
 and $Z+$jets (dashed) 
after the cut Eq.(\ref{basiccut}) and Eq.(\ref{ptcut}). 
BG events are rescaled to 50 fb$^{-1}$.}
\label{toppartner_label5}
\end{figure}

\begin{figure}
\includegraphics[scale=0.5]{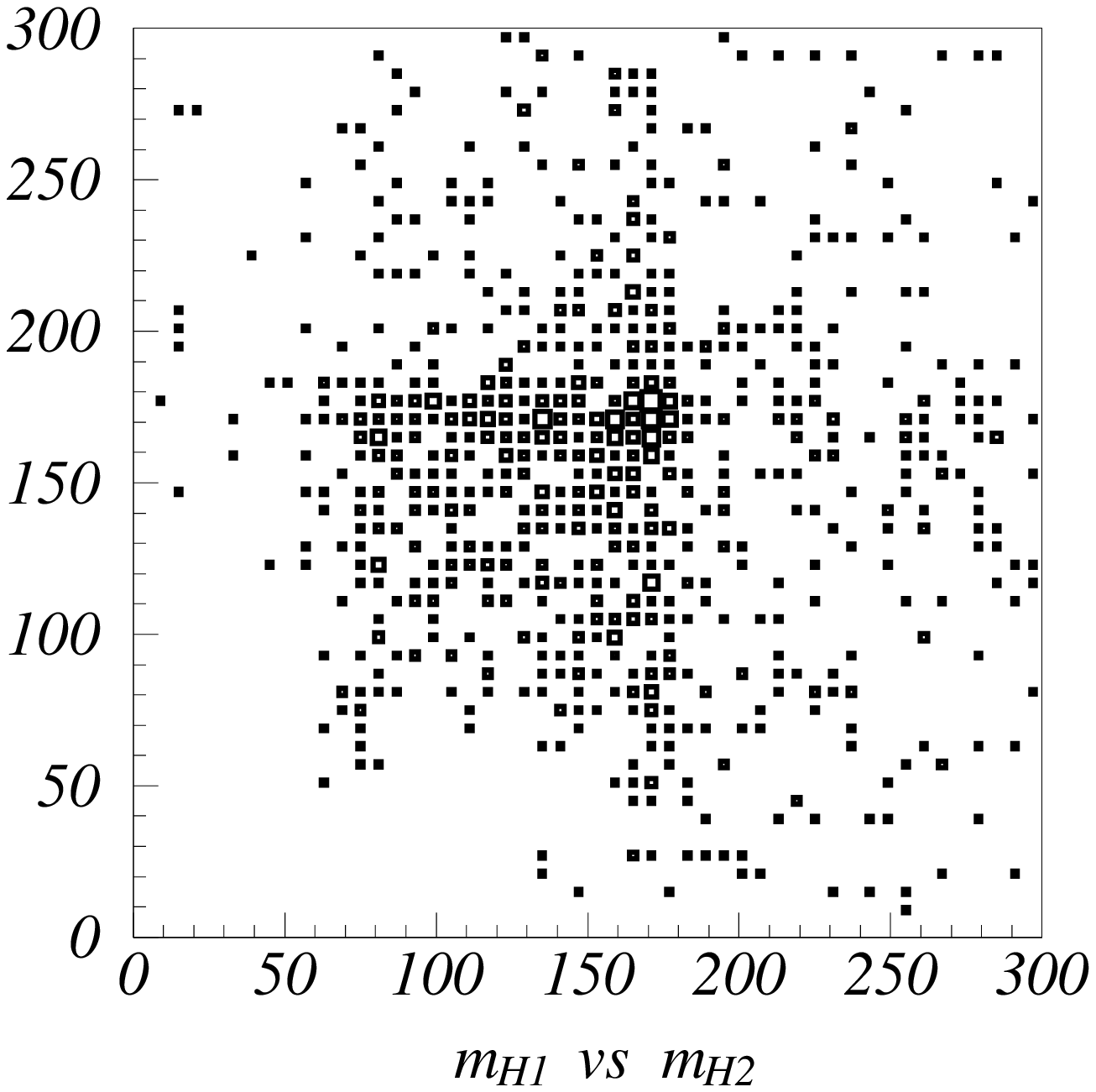}
\includegraphics[scale=0.5]{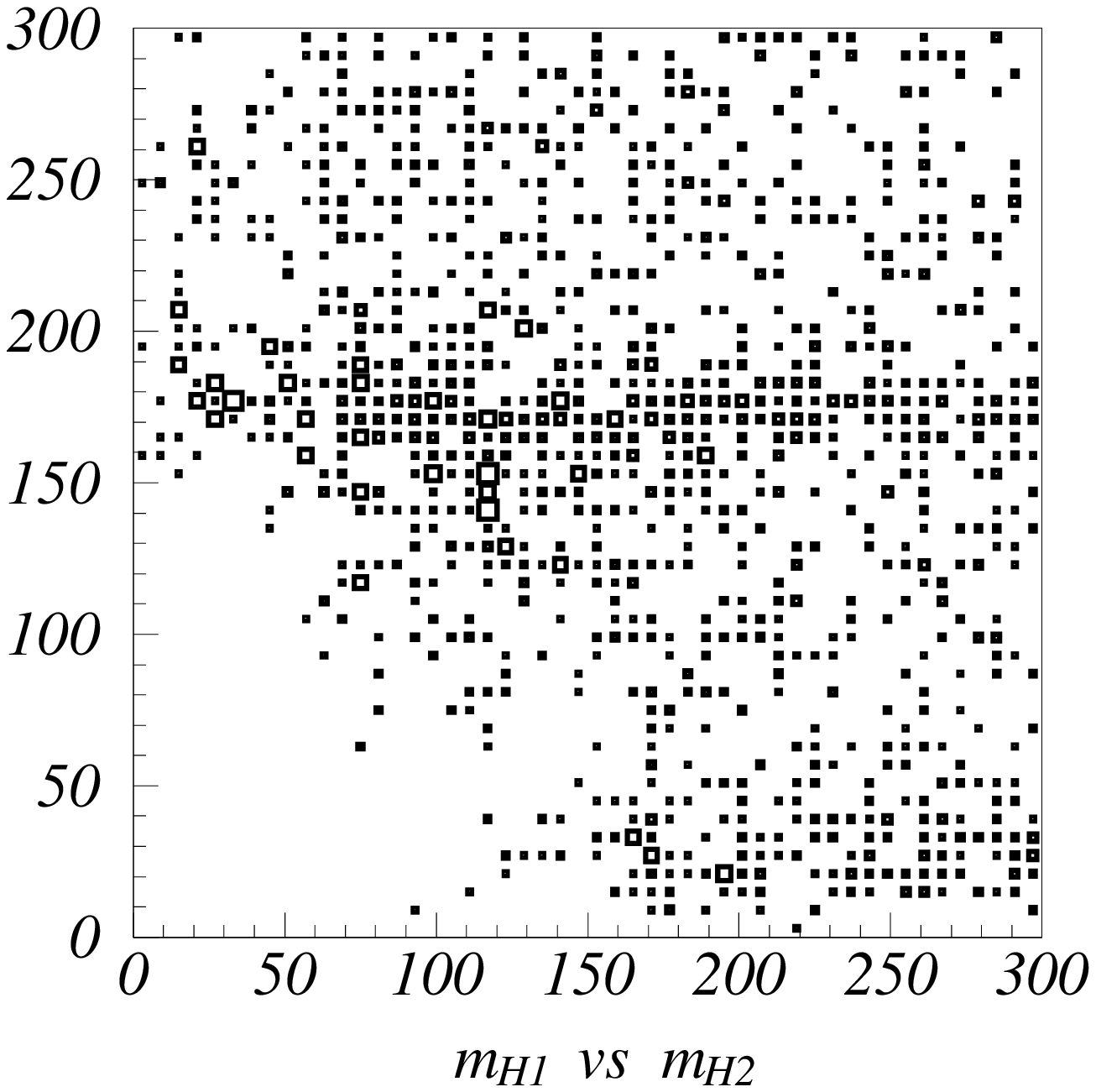}
\caption{$m_{P_{H_1}}$(vertical) vs. $m_{P_{H_2}}$(horizontal) distributions
for $T_-\overline{T}_-$ (the left figure) and 
for $t\bar{t}$ (the right figure).
The units of axes are GeV.
}
\label{toppartner_label6}
\end{figure}

After imposing these cuts, distributions of the invariant masses of the  hemisphere momenta
($m_{P_{H_i}}\equiv \sqrt{P_{H_i}^2}$) for the $T_-\overline{T}_-$ and the Standard Model background
events are shown in Fig.~\ref{toppartner_label5}.
We can see peaks at the top mass  both for the  $T_-\overline{T}_-$ and $t\bar{t}+$jets events 
in $m_{P_{H1}}$ distributions (Fig.~\ref{toppartner_label5}a, and \ref{toppartner_label5}c).
On the other hand, such a peak is not  seen  in the $m_{P_{H2}}$ distribution
 for $t\bar{t}+$jets events (Fig.~\ref{toppartner_label5}d).
This is because at least one of the two  tops 
should decay leptonicaly to give large 
$E\!\!\!/_T$. 
We also plot the distribution for $Z+$  jets and $W+ $ jets with dashed and dotted lines respectively.  
We do not  see any structure in the hemisphere mass
 distributions.
Two dimensional scattering plots in $m_{P_{H_1}}$ vs. $m_{P_{H_2}}$ 
plane for the signal and $t\bar{t}+$jets events are also shown in Fig.~\ref{toppartner_label6},
which show the clear  difference between them.

We can reduce the events from  $Z+$ jets, $W+$ jets and $t\bar{t}+$ jets 
with hemisphere mass cuts. In Table \ref{toppartner_label3}, the column $m_{P_{H1}}\sim m_t$
 shows the number of signal and background events after requiring
  $150$~GeV $<m_{P_{H1}}<190$~GeV. 
 The number of events of $t\bar{t}+$ jets ($Z+$ jets, $W+$ jets) 
 decreases by approximately $ 1/6\, (1/13, 1/17)$ after the cut.
 In the second hemisphere, only the signal distribution (Fig. 1b) has a peak 
 and background distributions (Fig. 1d) are flat.  

We do not simulate QCD background in this paper.
The magnitude of the QCD background for SUSY 0-lepton channel
is approximately the same as that of the $t\bar{t}+$ jets background \cite{SUSY08deJongtalk}.
The hemisphere mass distribution of QCD background should be similar to that of $Z+$ jets or $W+$ jets background.
Therefore the contribution of QCD background after the hemisphere mass cuts may be approximately 
the same as that of $Z+$ jets or $W+$ jets background.

The column \lq\lq both" %" 
 shows the number  of events after requiring the cut that
 both $m_{P_{H1}}$ and $m_{P_{H2}}$ are consistent with $m_t$. 
 The number of background becomes  small by a factor of $\sim 1/100$.
  However, the cut also reduces the signal 
  events by a factor of $1/10$. 
This is reasonable because we have the minimum jet energy cut for the hemisphere reconstruction
($p_T>30$~GeV), and some of top decay products may not contribute to the hemisphere momentum.  
Additionally, in the case that a $b$-quark decays semi-leptonically, 
the hemisphere momentum and the invariant mass are also reduced.
Maybe the approach in Ref.~\cite{Butterworth:2008iy} improves the mass resolution further. 
  At this point, the background still dominate the signal,  $S/N\sim 1/3$.  
If the sideband events 
 can be used to estimate the background distribution, the 
 significance of the signal events 
 goes beyond 5 sigma.

\begin{figure}[h!]
\includegraphics[scale=0.6]{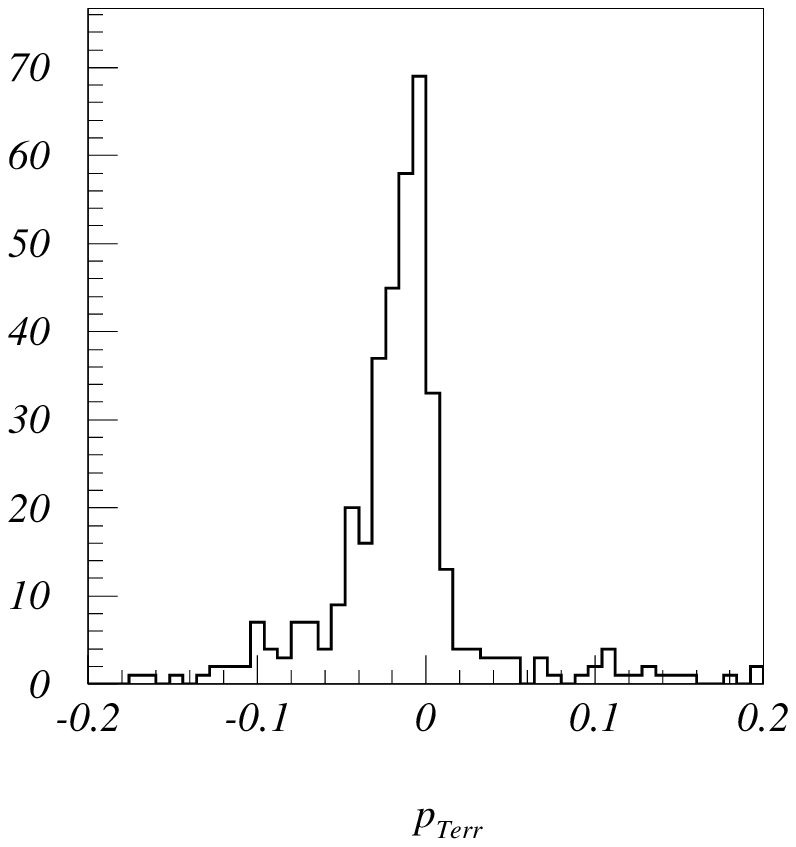}
\includegraphics[scale=0.6]{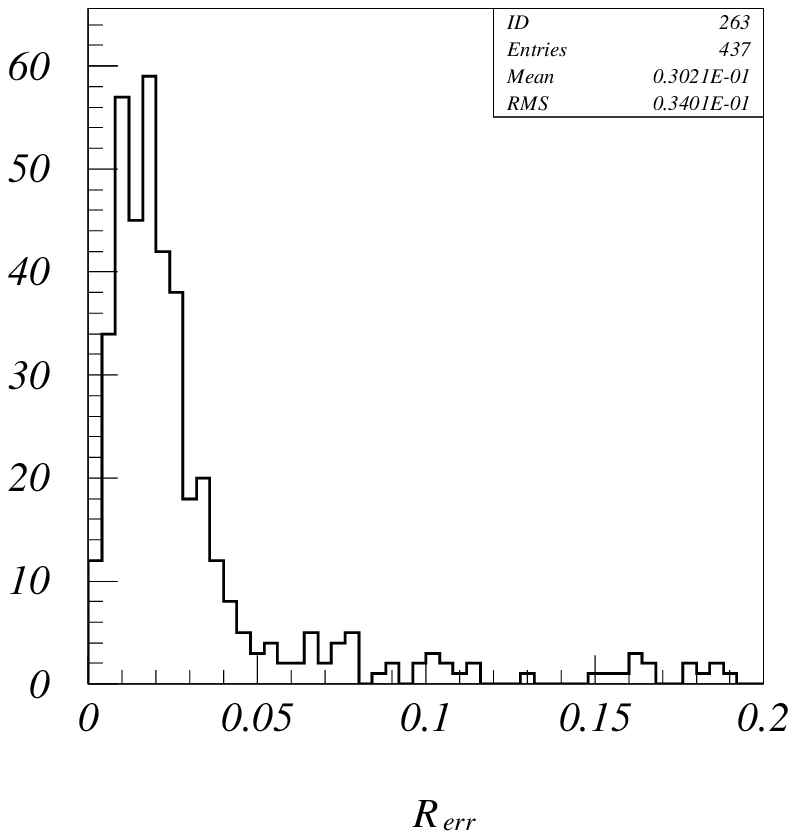}
\caption{$\Delta p_T/p_{T,H}$ and $\Delta R$  for $p_{H_1}$.}
\label{cbpartonmatch}
\end{figure}

To verify whether a momentum of a hemisphere whose mass is consistent with $m_t$ correctly matches a top momentum, 
we compare the $P_{H_1}$
with momentum of top partons $p_{\rm part}$.
In Fig.~\ref{cbpartonmatch}, we show the distributions of the $\Delta p_T \equiv p_{T,H_1} - p_{T,{\rm part}}$
and $\Delta R\equiv \sqrt{(\Delta \eta)^2 + (\Delta \phi)^2}$
 to see the difference between the  two momenta.
Here, $\Delta R$ and $\Delta p_T$ is defined for one of the two top partons 
that gives smaller $\Delta R$. 
The $\Delta p_T/p_{T,{H_1}}$ distribution has peak near 0, and $\Delta R < 0.05$
for most of the events. 
We conclude that a momentum of a hemisphere with $m_{P_H} \sim m_t$
may be considered as a momentum of a top partons.

\subsection{Measurement of the end point of $m_{T2}$}
\label{subsec_distribution}
We now show that top partner mass ($m_{T_-}$) can be measured 
using the endpoint of 
the distribution of the Cambridge $m_{T2}$ variable 
\cite{Barr:2003rg} for $t\bar{t}$ + $E\slush_T$ system 
if we know the LTP mass ($m_{A_H}$).
First, only the signal distribution is
considered and after that we will show 
that the background event does not contribute
to the events near the endpoint
and can be neglected for the determination of the endpoint.
It turns to be the best discrimination 
between the signal and SM backgrounds.

This variable is defined in the event  
$ \zeta \zeta^\prime \to (a\alpha) (b\beta)$,
where $\zeta$ and $\zeta^\prime$ have the same masses $m_\zeta$,
$a$ and $b$ are visible objects, 
and $\alpha$ and $\beta$ are invisible  particles 
with the same mass $M$. 
In such a event, $m_{T2}$ variable is defined as follows,
\begin{eqnarray}
\!\!\!\!
m_{T2}(\mathbf{p}_T^{a},\mathbf{p}_T^{b},\mathbf{p} {\!\!\!/}_T;M_{\rm test}) 
\equiv 
\min_{\mathbf{p} {\!\!\!/}_T^{\alpha} + \mathbf{p} {\!\!\!/}_T^{\beta} = \mathbf{p} {\!\!\!/}_T}
\left[
\max \left\{
m_T(\mathbf{p}_T^a,\mathbf{p} {\!\!\!/}_T^{\alpha};M_{\rm test}),
m_T(\mathbf{p}_T^b,\mathbf{p} {\!\!\!/}_T^{\beta};M_{\rm test}) 
\right\}
\right].
\end{eqnarray}

Here, $M_{\rm test}$ is an arbitrary chosen test mass and the transverse mass $m_T$ is defined as follows,
\begin{eqnarray}
m_T^2(\mathbf{p}_T^a,\mathbf{p} \slush_T^{\alpha};M_{\rm test})
\equiv m_a^2 + M_{\rm test}^2
+ 2 
\left[ E_T^a E \slush_T^\alpha -  \mathbf{p}_T ^{a}\mathbf{p} \slush_T^{\alpha}\right].
\end{eqnarray}
It is important that the following condition is satisfied in the case $M_{\rm test}=M$:
\begin{eqnarray}
 m_{T2}(M) \le m_\zeta.
\end{eqnarray}
Thus $m_{\zeta}$ can be extracted with measuring the upper endpoint of the $m_{T2}(M)$ distribution 
($m_{T2}^{\rm max}(M)$) in the case that the true $M$ is known.

In the case the true $M$ is not known, we can calculate $m_{T2}(M_{\rm test})$ for an arbitrary 
test mass $M_{\rm test}$.
For each test mass $M_{\rm test}$, we can measure $m_{T2}^{\rm max}(M_{\rm test})$.
The end point is expressed in terms of the following equation for 
the case that the masses of visible systems are the same ($m_{\rm vis}$)
and there is neither initial nor final state radiation \cite{Cho:2007dh},
\begin{eqnarray}
m_{T2}^{\rm max}(M_{\rm test})&=&\frac{m_{\zeta}^2 + m_{\rm vis}^2 - M^2}{2 m_{\zeta}}
+ \sqrt{\left( \frac{m_{\zeta}^2 + m_{\rm vis}^2 - M^2}{2 m_{\zeta}} \right)^2 + M_{\rm test}^2 - m_{\rm vis}^2}.
\label{mt2noisr}
\end{eqnarray}
The endpoint contains the information on a combination of the 
relevant masses $m_\zeta$ and $M$. 
For a particle which undergoes more complicated decay process,
$m_{\rm vis}$ can be various values.
Therefore one can obtain more than two independent information on the masses.
Practically, one can extract the true mass $M$ 
from a kink structure of the $m_{T2}$ endpoint as a function of $M_{\rm test}$ \cite{Cho:2007dh}.

\begin{figure}
\includegraphics[scale=0.5]{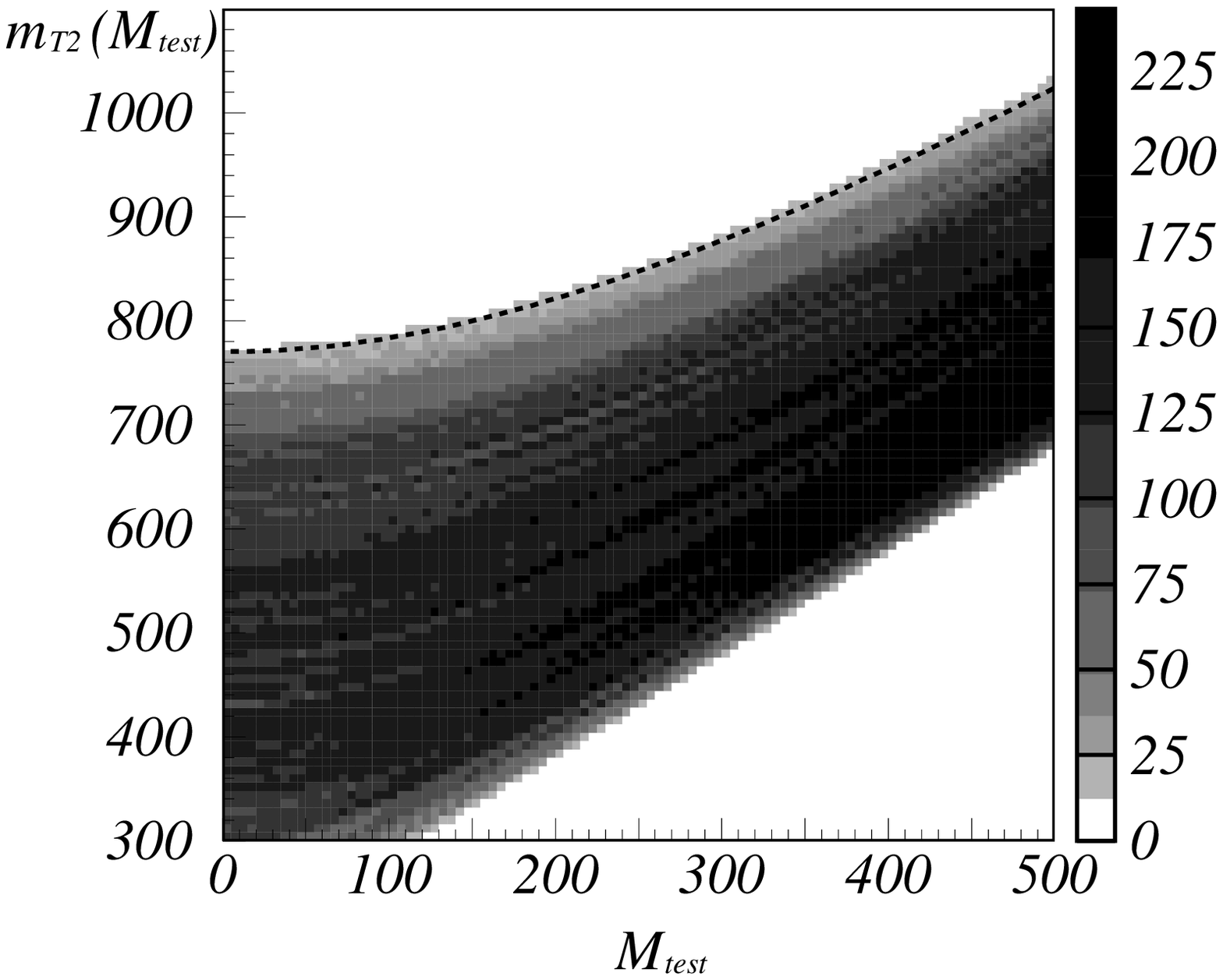}
\hfill
\includegraphics[scale=0.5]{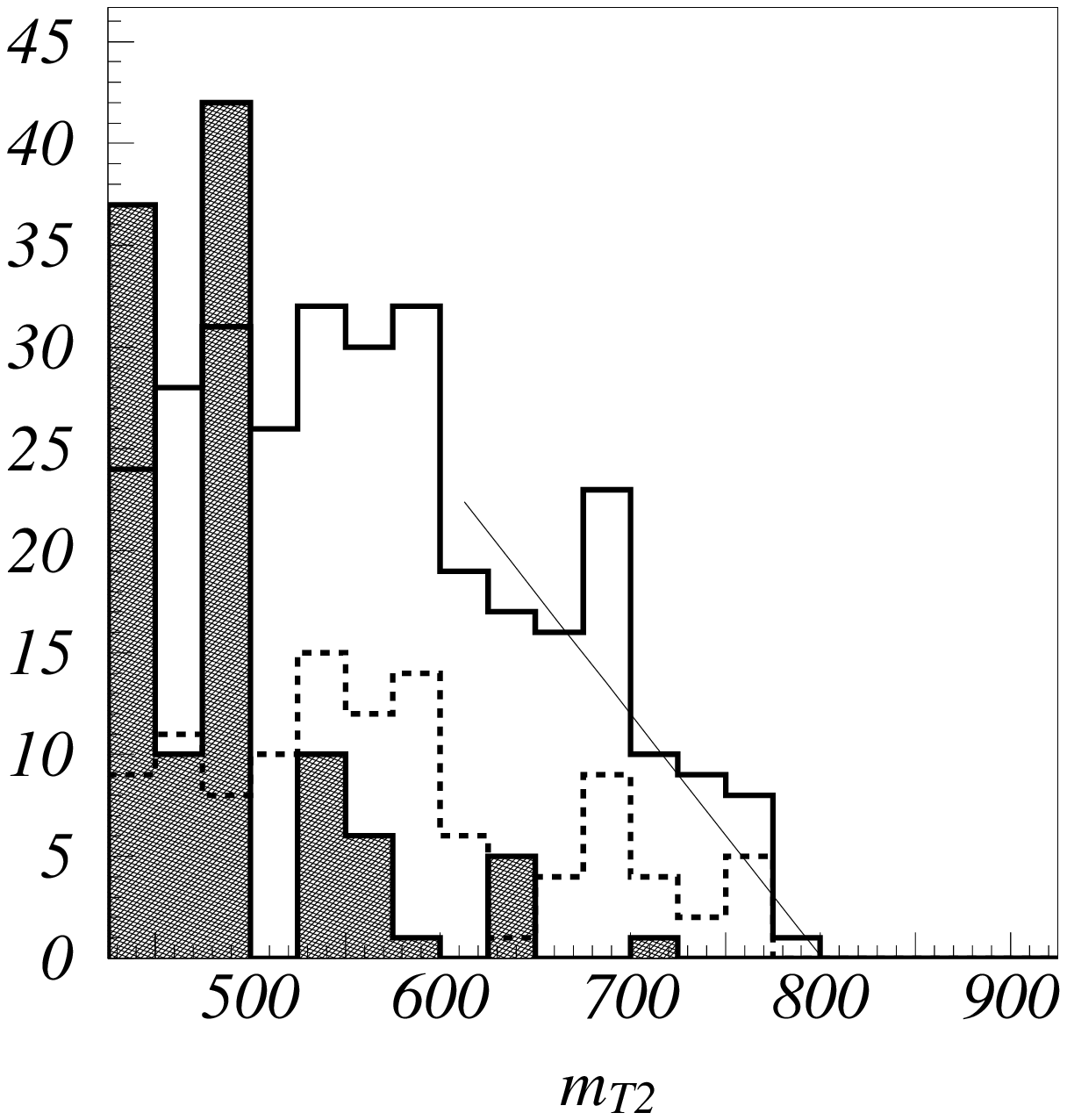}

\caption{left) $m_{T2}(M_{\rm test}) - M_{\rm test}$ 2dim. scattering plot at the parton level.
the dashed line is the line defined by the theoretical $m_{T2}^{\rm max}(M_{\rm test})$ 
description with no transverse momentum limit,
right)
The $m_{T2}$ distributions
for the nominal mass  $m_{A_H}=151.8$~GeV for the \lq\lq or" (solid line). 
The dashed line shows the contribution from the events survived after the \lq\lq both" cut.
The endpoint is 801.0 $\pm$ 9.4~GeV, and this value is consistent with $m_{T_-}=800$~GeV. 
}
\label{toppartner_label4}
\end{figure}

For our case, visible particles are two top quarks, and 
invisible particles are two $A_H$'s. 
It is, therefore, not possible to determine top partner mass itself 
from the kink method because $m_{\rm vis}$ is always $m_t$ 
and not to be various values.
We show the $m_{T2}(M_{\rm test}) - M_{\rm test}$ 2-dimensional scattering plot in the Fig \ref{toppartner_label4}.
The dashed line is the line defined by eq.(\ref{mt2noisr}) substituted with the nominal values.
The test mass dependence of the endpoint $m_{T2}^{\rm max}(M_{\rm test})$
is well described by eq.(\ref{mt2noisr}) and
no detectable kink structure can be seen.
Eventually, we can measure only a combination of masses;
\begin{eqnarray}
\frac{m_{T_-}^2 + m_{t}^2 - m_{A_H}^2}{2 m_{T_-}}.
\end{eqnarray}

In the case that 
a system of pair-produced particles has a net transverse momentum,
the $m_{T2}^{\rm max}(M_{\rm test})$  changes greater 
than the eq.(\ref{mt2noisr}) for all $M_{\rm test}$ 
but for $M_{\rm test}=m_{A_H}$,
therefore a kink structure might be seen \cite{Barr:2007hy}.
However, $T_- \overline{T}_-$ system 
generally has a small net $|p_T| \sim O(100)\,$GeV in average,
a kink structure is not seen in our case 
even at parton level.
Then we cannot measure the $m_{T_-}$ unless the $m_{A_H}$ is known for our case.
The $m_{A_H}$ may be determined if productions of the other 
$T$-odd particles are observed.
Alternatively, if we assume the thermal relic density of $A_H$ 
is consistent with the dark matter density in our universe, $m_{A_H}$ is 
related to the Higgs mass so that it is determined with two fold ambiguities \cite{Asano:2006nr}. 

Next,
we show that $m_{T_-}$ can be measured using reconstructed tops at jet level
assuming $m_{A_H}$ is known.
Ideally we may regard hemisphere momenta as top momenta
if both hemisphere masses satisfy the condition $150{\rm ~GeV} < m_{P_{H}} < 190{\rm ~GeV}$.
There is, however, not enough number of events left under these cuts 
as mentioned in the previous subsection.
Therefore in the following, we  apply the cuts 
that one of the  hemisphere masses $m_{P_H}$ 
satisfies $150$~GeV$<m_{P_H}<$190~GeV 
while the other satisfy $50$~GeV $<m_{P_H}<190$~GeV  (the column 
\lq\lq or" %"
in the table) 
and regard the hemisphere momenta as top momenta.

The endpoint of the $m_{T2}$ distribution does not change under the relaxed cut
\lq\lq or". %" 
because $m_{T2}$ is an increasing function of visible masses.
Additional sources of missing momentum (such as neutrinos) do not affect the endpoint either.
It is easy to understand this as follows,
a system of a LTP and the other sources of missing momentum
can be regarded as an invisible pseudo-particle.
The invisible pseudo-particle's invariant mass 
is always larger than the LTP mass 
($m_{\rm invisible} \equiv \sqrt{(p_{\rm LTP} + p_{\rm other})^2} \ge m_{A_H}$),
 and $m_{T2}(m_{\rm invisible})$ is nevertheless smaller than $m_{\zeta}$
because the system comes from $\zeta \zeta^\prime$ pair production.
On the other hand, $m_{T2}(M_{\rm test})$ is a monotonically increasing function of $M_{\rm test}$, 
therefore, $m_{T2}(m_{A_H}) \le m_{T2}(m_{\rm invisible}) \le m_\zeta$ is satisfied.

The $m_{T2}$ distributions for the nominal value $m_{A_H}=151.8$~GeV
are shown in Fig.~\ref{toppartner_label4}(right).
The distribution after the \lq\lq or" %"
cut for the signal events are shown in the solid line.
We fit the distribution near the endpoint by a linear function and
obtain $m_{T2}^{\rm max}=801.0 \pm 9.4$~GeV.
This value is consistent with the nominal value 
$m_{T_-}=800$~GeV.\footnote{In Fig.\ref{toppartner_label4},
we do not include the effects of jet energy smearing. 
In Appendix \ref{smearing}, we discuss the point.}
The fact supports validity of the relaxed cut (\lq\lq or" cut) 
in determination of the $m_{T2}$ endpoint.
The contribution from the events survived 
after \lq\lq both" cut is shown in the dashed line
and there are a few statistics.

The distribution for the SM backgrounds is also shown in a dark histogram,
and they have lower $m_{T2}$ values.
After imposing the cut $m_{T2} > 350$GeV, the SM background is significantly reduced 
but the signal is not reduced as in Table \ref{toppartner_label3}.
Moreover, after imposing the cut $m_{T2} > 500$GeV, the SM background becomes negligible.
Therefore we can neglect them
to fit the endpoint $\sim 800$GeV in the present case.

In a case that $m_{T_-}$ is lighter, for example $m_{T_-}=500$GeV,
a top quark momentum arising from a top partner is approximately 200GeV ($m_{A_H}=150$GeV is assumed).
It is boosted enough and the decay products distribute within $\Delta R < 1.5$ \cite{boostedtop},
then the hemisphere analysis may work well.
And the $\sigma(T_- \overline{T}_-)$ is ten times larger than the case of $m_{T_-}=800$GeV \cite{Belyaev:2006jh}.
Therefore it is possible to measure the endpoint by the method we proposed above\footnote{
For discovery, we can also use lepton channels.
In Ref. \cite{Han:2008gy}, it is found that $T_-$ can be discovered
in the case of $\Delta M_{TA}=m_{T_-} - m_{A_H} \sim 250$GeV.}. 

\section{Comparison among jet reconstruction algorithms}
\label{toppartner_label2c}
\subsection{Jet reconstruction algorithms}
\label{toppartner_label2}
We now study a dependence of the  signal distributions on 
jet reconstructing algorithms.
The reason to study different jet reconstruction algorithms 
is as follows.   Note that  we need to study the 
jet system arising from a boosted top quark. In the 
rest frame of a top partner, a top quark momentum arising from 
top partner decay is expressed as 
\begin{equation}
p_t=\frac{m_{T_-}}{2}
 \sqrt{1-2\frac{m_t^2+m_{A_H}^2}{m_{T_-}^2} + \frac{(m_t^2-m_{A_H}^2)^2}{m_{T_-}^4}} \sim 365{\,\rm GeV},
\end{equation}
therefore, typical $p_T$ of a top quark is above  300~GeV. 
The jet angle separation is  of the order of 
$\theta \sim m_T/p_T$. 
If decay products of a top are aligned in the direction of the top momentum, the angle is even smaller. 
It is important to choose the algorithm that 
gives the best result in such a situation.
Four algorithms (Snowmass cone, kt,Cambridge, SISCone) are used in the following analyses.

\subsubsection{Cone algorithms}
We take two cone-type algorithms. The first one is \lq\lq Snowmass cone",%"
which is  a simple algorithm implemented in {\tt AcerDET1.0}. 
It defines a list of jets as follows,
\begin{center}
\fbox{
\begin{minipage}{15cm}
\begin{enumerate}
\item 
Find the particle $i$ which has the maximum $E$ in all particles,
and take it as a seed.
If $E_i$ is less than some threshold $E_{\rm th}$ then the process is finished.
\item 
Sum the four-momenta of the particles in the circle of whose center and radius are
$(\eta_{\rm seed},\phi_{\rm seed})$ and $R$ respectively.
Define the four-momentum
as $p_{\rm cone}$ and redefine it as a new seed.
\item 
Repeat Step 2 until $p_{\rm cone}$ is converged.\footnote{{\tt AcerDET} jet finding algorithm skip this iteration.}
\item 
Remove the constituents of the cone from the particle list
and repeat from Step 1.
\end{enumerate}
\end{minipage}
}
\end{center}
In this paper,
we take cell momenta as massless particle momenta.
As one can easily see from the algorithm, the highest $p_T$ jet 
in a region takes all activities within $R(<0.4)$ even if there are sub-dominant
activities nearby $R \sim 0.4$.  As we will see later, jet-parton energy matching is worse 
than the other algorithms. 

The second one is SISCone \cite{Salam:2007xv}.
This algorithm is a seedless cone search algorithm.
It defines a list of jets as follows,

\begin{center}
\fbox{
\begin{minipage}{15cm}
\begin{enumerate}
\item 
Find all \lq\lq stable" cones seedlessly and calculate four-momenta of these cones.
Here, a \lq\lq stable" cone is defined with a set of particles satisfying the following condition,
\begin{eqnarray}
\sum_{d(p_i,p_{\rm cone}) < R} p_i = p_{\rm cone}
\end{eqnarray}
$d(p_i,p_j)$ is the distance in the $\eta-\phi$ plane.
\item 
Remove cones which have less energy than some threshold $E_{\rm th}$
from the cone list.
\item 
If there are overlapping cones,
determine to split or merge according to the overlap parameter $f$.
Namely if the fraction of overlapping activities of the two jets by
the smaller jets is larger than $f$ two jets are merged,
otherwise split the overlapping activities into the two 
jets.
And update the cone list.
If there is a cone which is not overlapping with other cones,
remove it from the cone list and add it to the jet list.
\item 
Repeat Step 3 until there is no cone in the cone list.
\end{enumerate}
\end{minipage}
}
\end{center}
The reconstruction algorithm is infrared safe, because the reconstruction 
does not relay on the highest $p_T$ cell in a cell list. The number of 
reconstructed jets depends sensitively on the overlap parameter $f$.
If we set $f$ smaller, the algorithm tends to merge jets.
For our choice $f=0.75$, which is the default value, jets from a top quark tend to 
be merged and efficiency of resolving the three jets in a hemisphere is 
rather low compared with the other algorithms.

\subsubsection{Clustering algorithms}
The other category of jet finding algorithms is 
a clustering algorithm.
A typical algorithm in this category, 
the kt algorithm \cite{Catani:1991hj,Catani:1993hr,Ellis:1993tq,Cacciari:2005hq}
is defined as follows:

\begin{center}
\fbox{
\begin{minipage}{15cm}
\begin{enumerate}
\item 
Work out the $k_t$ distance $d_{ij}$ for each pair of particles  with 
momentum $k_{i},k_{j}$ 
and  $d_{iB}$ for each particle $i$.
\begin{eqnarray}
 d_{ij}\equiv \min(k^2_{ti},k^2_{tj}) \frac{R^2_{ij}}{R^2},\ \ \ \ \ 
d_{iB} \equiv k^2_{ti},\ \ \ 
R^2_{ij}\equiv (\Delta \eta)_{ij}^2 + (\Delta \phi)_{ij}^2.
\label{ktdef}
\end{eqnarray}
\item 
Find the minimum $d_{\rm min}$ of all the $d_{ij},d_{iB}$.
If $d_{\rm min}$ is a  $d_{ij}$, merge the particles $i$ and $j$
into a single particle by summing their four-momenta.
If  the $d_{\rm min}$ is a $d_{iB}$ then
regard the particle $i$ as a final jet and remove it from the list.
\item 
Repeat from Step 1 until no particles are left.
\end{enumerate}
\end{minipage}
}
\end{center}
Cambridge algorithm \cite{Dokshitzer:1997in,Bentvelsen:1998ug} is similar to the kt algorithm
but definition of $d_{ij}$ and $d_{iB}$ is modified as follows:
\begin{center}
\fbox{
\begin{minipage}{15cm}
\begin{equation}
d_{ij} \equiv \frac{R^2_{ij}}{R^2},\ \ \ \ \ \ \ 
d_{iB} \equiv 1.
\end{equation}
\end{minipage}
}
\end{center}

\subsection{Hemisphere invariant mass distributions}
\label{toppartner_label7}

In Table \ref{toppartner_label8}, we show the numbers of signal events 
for the four jet algorithms
after the same hemisphere mass cuts as in Table \ref{toppartner_label3}.
{\tt AcerDET} has an option to rescale jet energy,
the results for the Snowmass cone algorithm are given
with jet calibration.
The scale factor is determined so that
an invariant mass distribution of the two jets from $W$ has the 
peak consistent with $m_W$.

\begin{table}[htbp]
\centering
\footnotesize
\begin{tabular}{|c||r|r|r|r||r|}
\hline 
 For $ T_-\overline{T}_-$ events     &
$m_{P_{H_1}}\sim m_t$ &$m_{P_{H_2}}\sim m_t$ &$\ \ $ both$\ \ $ & $\ \ $or$\ \ $ &\scriptsize{ $m_{T2}$ endpoint} \cr
\hline 
Snowmass cone (calibrated)& 375 & 442 & 143 & 363 & $795$~GeV\cr
  kt                      & 420 & 415 & 135 & 411 & $797$~GeV\cr
  Cambridge               & 404 & 396 & 130 & 398 & $801$~GeV\cr
  SISCone                 & 425 & 385 & 137 & 396 & $796$~GeV\cr
\hline
\end{tabular}
\caption{Summary of the cuts for various jet finding algorithms.
The cuts are the same as in Table \ref{toppartner_label3}.}
\label{toppartner_label8}
\end{table}

We obtained 375, 420, 404, 425 events  after the $m_{P_{H_1}} \sim m_t$ cut 
for the Snowmass cone, kt, Cambridge, SISCone respectively\footnote{As we mentioned already, the 
Snowmass cone algorithm in {\tt AcerDET} ignores jet invariant masses, therefore 
jet energy calibration should increase a jet energy to 
compensate the missing jet mass. 
The calibration also compensate average energy of particles 
that fall outside  jet  cones. For our case,  several jets 
go collinear, and the particles outside the cone often fall into  other 
jet cones, leading  overestimate of the jet energies.}.

\begin{figure}[htbp]
\begin{tabular}{cc}
\begin{minipage}[t]{14cm}
\includegraphics[scale=0.6]{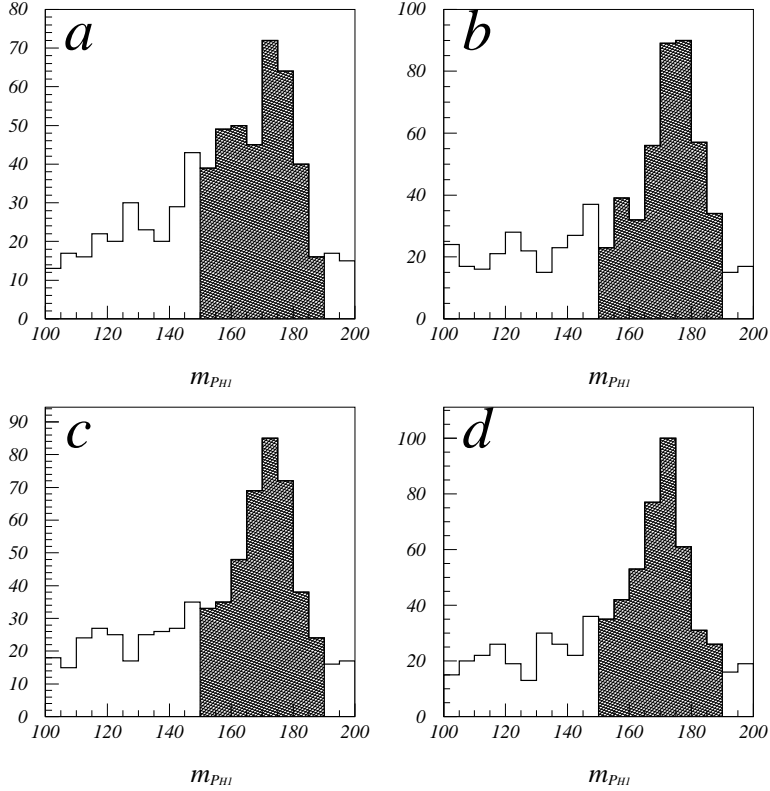}
\caption{The distributions of  $m_{P_{H_1}}$ for the $T_-\overline{T}_-$ events
for a) Snowmass cone ({\tt AcerDET}) , b) kt, c) Cambridge and d) SISCone.
}
\label{toppartner_label9}
\end{minipage}
\end{tabular}
\end{figure}

The distributions of $m_{P_{H_1}}$ for four algorithms
are shown in Fig.~\ref{toppartner_label9}.
The shaded regions denote the region satisfying $150{\rm ~GeV} < m_{P_H} < 190{\rm ~GeV}$.
The kt, Cambridge and SISCone show nice resolutions in top mass.
The peak for the Snowmass cone algorithm is dull and has broad tail.
This is because  {\tt AcerDET} takes massless jets.  
This is rather an artificial difference as it is straight forward 
to define non-zero 
jet masses from calorimeter information. If such  jet definitions are feasible 
at the LHC environment, the reconstruction efficiency may 
 be increased significantly although we have not simulate
the effect of mis-measurement of calorimeter energy. 
We regard the efficiency in the Snowmass cone as a conservative 
estimate. 
Fortunately,
the endpoints of $m_{T2}$ distributions are
rather insensitive to the reconstruction algorithm.  We find that they are
$795$~GeV (Snowmass cone),
$794$~GeV (kt),
$801$~GeV (Cambridge),
$796$~GeV (SISCone) with the statistical errors of the order of 10 GeV.
The difference among the algorithms is not essential at this point.

\subsection{Parton-jet  matching}

\begin{figure}[htbp]
\includegraphics[scale=0.45]{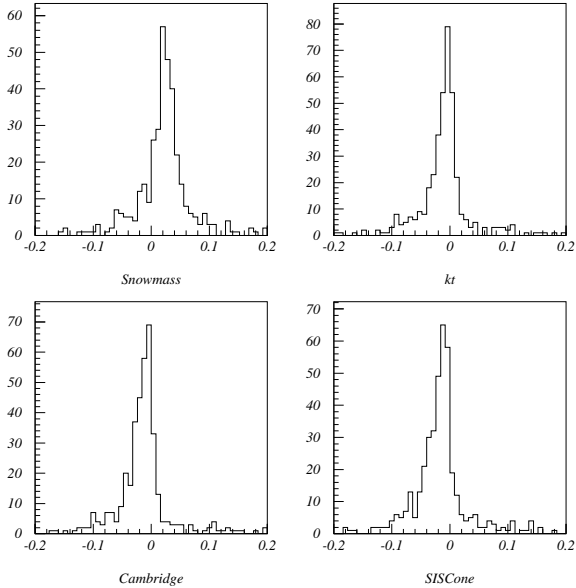}
\caption{$\Delta p_T/p_{T,H_1}$  distributions of hemisphere $H_1$ 
for $T_-\overline{T}_-$ events.
The events with 150~GeV$< m_{P_{H_1}} <$ 190~GeV are selected.
}
\label{partonmatchp}
\end{figure}

Fig.~\ref{partonmatchp} shows deviation of a hemisphere momentum
from a true top parton momentum $\Delta P_T/P_{T,H_1}$ for the four algorithms.
We selected the signal events with 150~GeV$< m_{P_{H_1}}<$ 190~GeV.
The $\Delta P_T/P_{T,H_i}$ is mainly distributed within $\pm 5$\% region for all algorithms.
The $\Delta R$ for top parton and hemisphere momentum 
($=\sqrt{(\eta_{H}-\eta_{\rm top})^2+(\phi_{H}-\phi_{\rm top})^2}$) is mainly distributed 
less than 0.03 for all algorithms.
These agreements justify regarding a hemisphere momentum as a top momentum.
The peak position is larger than 0 by approximately 2\% for the Snowmass cone algorithm,
because jet energy calibration by {\tt AcerDET} 
leads over-estimate of the jet energy for collimated jets.
The jet energy calibration compensates the activities outside the jet cone,
but for the collimated jets they are taken into account by the other jets.
For the other algorithms, the peak position is less than 0 
due to semi-leptonic decays of $b$-quark. 

\begin{figure}[htbp]
\includegraphics[scale=0.45]{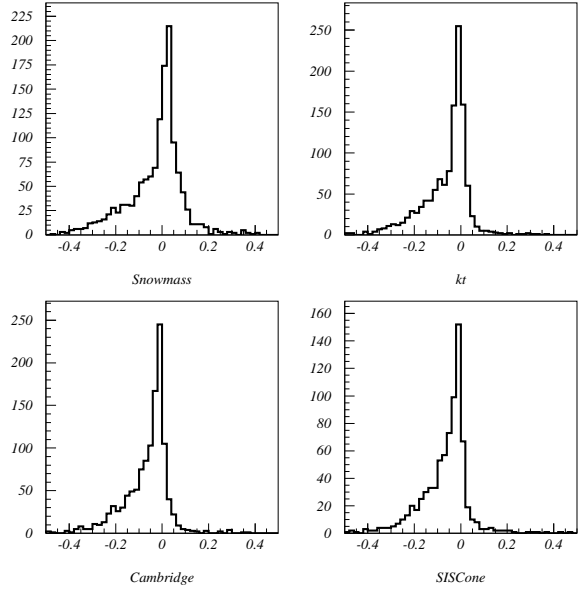}
\caption{$\Delta p_{T,b{\rm -jet}}/p_{T,b{\rm -part}}$  distributions for the 
selected hemispheres (150~GeV$< m_{P_H} <$ 190~GeV)
for the $T_-\overline{T}_-$ events.
}
\label{bpartonmatchp}
\end{figure}

We now look into the matching between a $b$-parton and a $b$-jet in a selected hemisphere.
The selected hemispheres satisfy the following conditions:
1) there are only three jets in the hemisphere and 150~GeV$<m_{jjj}<190$~GeV, 
and 2) at least one jet pair satisfies $|m_{jj} - m_W| < 20$~GeV and the other jet is $b$-tagged.
Here, we define a jet with $\Delta R_b <0.2$
as a $b$-jet
($\Delta R_b \equiv 
\sqrt{(\eta_{\rm jet}- \eta_{b{\rm -part}})^2 +
(\phi_{\rm jet}- \phi_{b{\rm -part}})^2})$.
In Fig.~\ref{bpartonmatchp}, the
$\Delta p_{T,b {\rm -jet}}/p_{T,b{\rm -part}}$ distributions are shown. 
The large tails found for $\Delta p_{T,b {\rm -jet}}/p_{T,b{\rm -part}}<0$ come from a $b$-parton decaying semi-leptonically.  
In addition, the distribution for the Snowmass cone algorithm shows a tail for 
$\Delta p_{T,b{\rm -jet}}/p_{T,b{\rm -part}} > 0$. 
This tail arises 
because a locally highest $p_T$ jet takes over all energy 
in $R=0.4$ cone 
because the jet finding algorithm starts from the highest $p_T$ clusters.
This feature cannot be improved with minor modifications of the algorithm.
For the SISCone algorithm, the number of the three jet events is significantly 
small compared with the others because the algorithm actively merges 
overlapping jets. We will discuss this point in the next subsection.
The algorithm is therefore not suitable for our analysis 
in the next section, in which we study the top spin 
dependence of $b$-jet distributions.

\subsection{Number of Jets distribution}

In this subsection we compare clustering algorithms with the SISCone
in terms of the number of jets in a hemisphere.
The numbers of jets inside a hemisphere 
with 150 GeV$<m_{P_H} <$ 190 GeV 
 are shown in the Table \ref{nofj4} and \ref{nofj2}.
Since kt algorithm behaves similar to Cambridge algorithm,
only those for the Cambridge and SISCone are shown.
The parameter $R$ for the clustering algorithms and for the SISCone have different meanings
and SISCone has additional parameter $f$ as explained in Sec.\ref{toppartner_label2}.
We investigate the distribution of the number of 
jets varying these parameters
($R=0.2,0.3,0.4$ and $f=0.5,0.75,0.9$ for SISCone).

In order to study the top decay distribution,
it is better to choose the parameters which give higher 3-jets acceptance.
With the Cambridge algorithm ($R=0.4$) 
511 events are classified into a group of 3-jets events, 
while with the SISCone ($R=0.4, f=0.75$) 324 events are classified into it
 although the total numbers of hemispheres 
with 150 GeV $< m_{P_H} <$ 190 GeV are approximately the same.
We can see $(R,f)=(0.3,0.9)$ or $(R,f)=(0.2,0.5)$ are optimal
to enhance the number of 3-jets events with the SISCone for our model point.
The distribution of numbers of jet at those parameters are similar 
to that with Cambridge ($R=0.4$).
For such a small $R$, however,
some activities are missed outside a jet cone leading 
worse parton-jet matching\footnote{For such a small $R$, detector granularity might not be
enough to resolve the jet}.
Moreover, we found Cambridge with $R=0.3$ also gives higher acceptance 
for 3-jets events than with $R=0.4$.
We do not find the parameter which
improves the results for SISCone over Cambridge
 by changing $R$ and $f$,
therefore, we use clustering algorithms for the further analysis.

Appropriate $R$ should be used depending on top $p_T$ to protect unnecessary merging.
Sub-jet analysis based on clustering algorithms might be useful 
 in such a case \cite{Butterworth:2008iy}.
We do not discuss these points any more because it is beyond the scope of this paper.

\begin{table}[htbp]
\centering
\footnotesize
\begin{tabular}{|c||r|r|r|r|}
\hline 
\multicolumn{1}{|c||} { $R=0.4$ } &\multicolumn{1}{c|} {Cambridge} & \multicolumn{3}{c|}{SISCone} \cr
\cline{3-5}
&\multicolumn{1}{c|}{}& $f=0.5$ &  $f=0.75$ & $f=0.9$ \cr
\hline
\hline
1 jet         &   9 & 103 &  58 &  41  \cr
2 jet         & 244 & 429 & 413 & 362  \cr
3 jet         & 511 & 295 & 324 & 362  \cr
4 jet or more &  36 &   5 &  15 &  23  \cr
\hline 
total        &  800 & 832 & 810 & 788  \cr
\hline
\end{tabular}
\caption{Summary of the number of jets in a hemisphere with $m_{P_H} \sim m_t$
for $R=0.4$.}
\label{nofj4}
\end{table}

\begin{table}[htbp]
\centering
\footnotesize
\begin{tabular}{|c||r|r|r|r||r|r|r|r|}
\hline 
\multicolumn{1}{|c||} {} &\multicolumn{1}{c|} {Cambridge} & \multicolumn{3}{c||}{SISCone $R=0.3$} 
&\multicolumn{1}{c|} {Cambridge} & \multicolumn{3}{c|}{SISCone $R=0.2$} \cr
\cline{3-5}\cline{7-9}
 &\multicolumn{1}{c|}{$R=0.3$}& $f=0.5$ &  $f=0.75$ & $f=0.9$ 
 &\multicolumn{1}{c|}{$R=0.2$}& $f=0.5$ &  $f=0.75$ & $f=0.9$ \cr
\hline 
\hline
1 jet         &   2 &  51 &  16 &  12 &   1 &  10 &   2 &   2  \cr
2 jet         & 138 & 335 & 294 & 235 &  47 & 135 & 106 &  84  \cr
3 jet         & 574 & 419 & 430 & 462 & 467 & 477 & 440 & 430  \cr
4 jet or more &  85 &  36 &  55 &  71 & 136 & 107 & 122 & 115  \cr
\hline
total        &  799 & 841 & 795 & 780 &  651 & 729 & 670 & 631  \cr
\hline
\end{tabular}
\caption{Summary of the number of jets in a hemisphere with $m_{P_H} \sim m_t$
 for $R=0.3$ and $0.2$.}
\label{nofj2}
\end{table}

\subsection{Effects of Underlying Events to the reconstruction}
\begin{figure}[htb]
\begin{minipage}{7.5cm}
\includegraphics[scale=0.45]{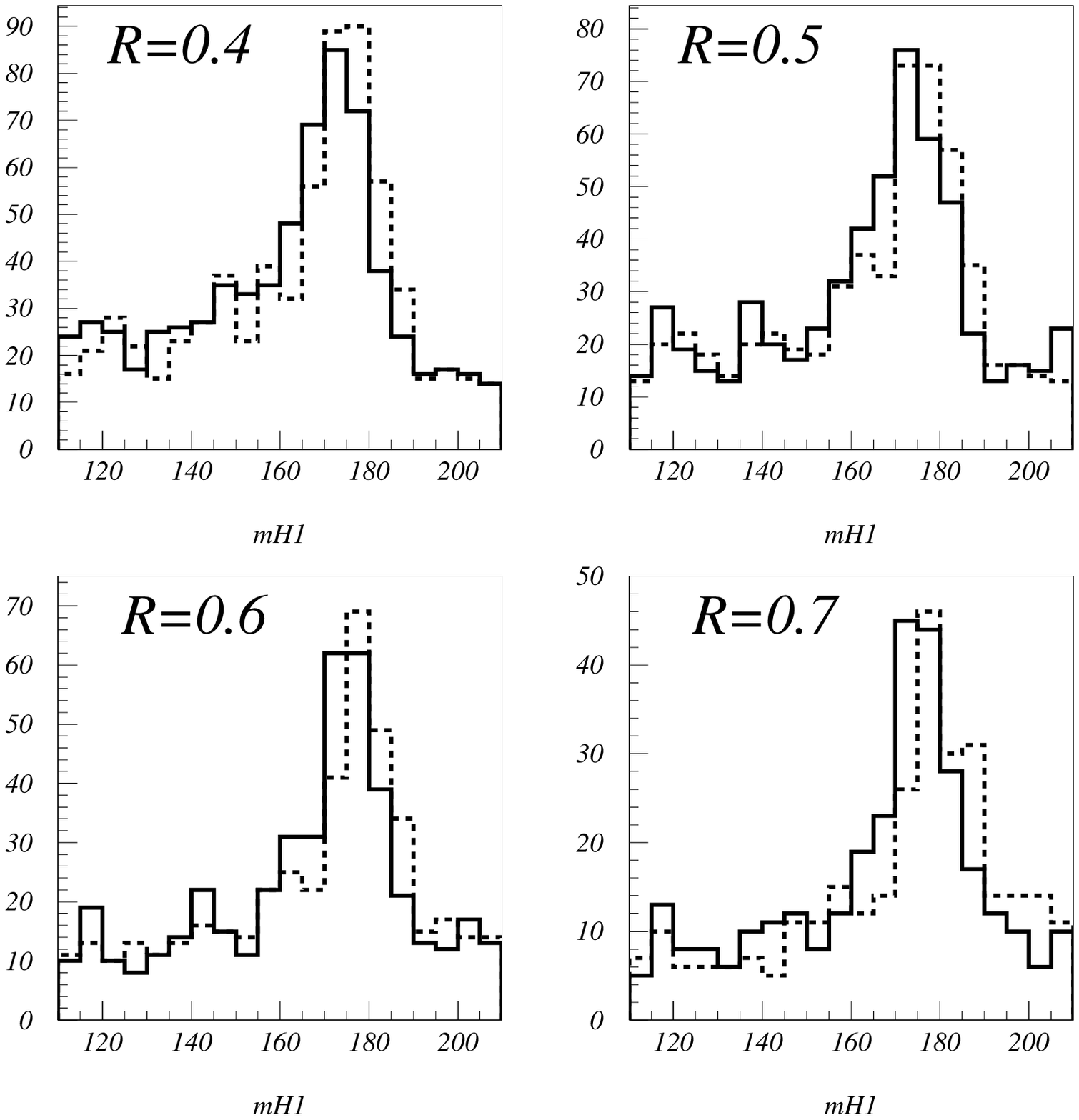}
Without underlying events
\end{minipage}
\hfill    
\begin{minipage}{7.5cm}
\includegraphics[scale=0.45]{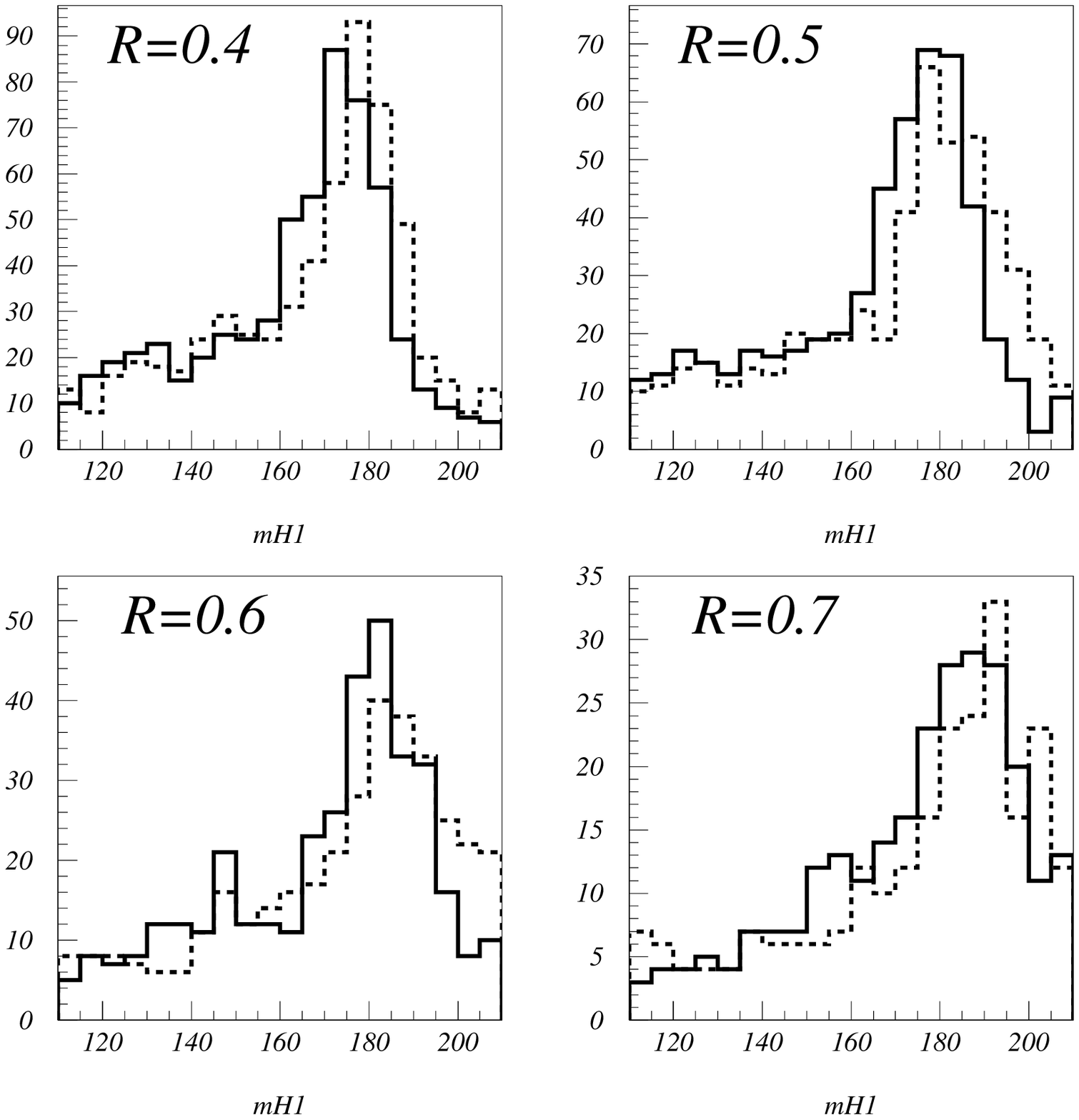}
With underlying events
\end{minipage}
\begin{flushleft}
\caption{The $m_{P_{H_1}}$ distributions in Cambridge (solid) and kt (dashed) 
algorithms without (left) and with (right) underlying events.
We take  $R=0.4 \sim 0.7$.}
\label{withoutjimmy}
\end{flushleft}
\end{figure}

So far we have discussed  the event distributions 
without underlying events.  
Underlying events come from 
the soft parton interactions which occur with a hard collision, and whose nature 
at the LHC has large theoretical uncertainty.
The top reconstruction efficiency may become worse with them, because 
the number of hit cells significantly increases with underlying events. 
We have generated the signal events with underlying events and multiple scattering 
using {\tt HERWIG6.5} + {\tt JIMMY} \cite{Butterworth:1996zw}. 
In Fig.~\ref{withoutjimmy} (left), we  show the distributions of $m_{P_{H_1}}$ 
for the kt (dashed) and Cambridge (solid) algorithms without underlying 
events for $R=0.4$, $0.5$, $0.6$, $0.7$.
The event selection cuts are the same as section \ref{toppartner_label1}.
We can see that the locations of the peaks increase as increasing $R$
for both the kt and Cambridge algorithms.
The shapes of distributions are similar for all $R$ as  top quarks are boosted enough 
so that the decay products are isolated from the other activities.
However the kt algorithm tends to give higher invariant mass than the Cambridge algorithm. 

In Fig.~\ref{withoutjimmy} (right), we  show the same distributions with underlying events.
We can see that the position of the peak for the kt algorithm is significantly larger than 
that for the Cambridge algorithm in all $R$ values. Even for $R=0.4$, the peak 
position is larger than 175~GeV for the kt algorithm.
The situation becomes worse for the kt algorithm as $R$ increases. 
The reconstruction 
efficiency is reduced significantly for $R\ge 0.6$. 
This is because 
the kt algorithm over-collects soft activities  which are far from the jet 
direction (large $R_{ij}$)
due to the factor $\min(k^2_{ti},k^2_{tj})$ in the definition of the  distance in Eq.(\ref{ktdef}), 
which is known as splash-in effects.
 On the other hand, the Cambridge distance measure does not have 
the factor, therefore, it is not too sensitive to the existence of the underlying events.
The effect of the underlying events can be safely neglected for $R\sim 0.4$.
Hence we take the Cambridge algorithm in 
Sec.~\ref{lht} and \ref{toppolarization}. 

\section{Top polarization effects}
\label{toppolarization}
In this section, we consider top polarization effects.
In the Littlest Higgs model with T-parity,
the Lagrangian relevant to a top partner decay is written as follows
\cite{Low:2004xc,Hubisz:2004ft,Hubisz:2005tx,Matsumoto:2006ws,Asano:2006nr}.
\begin{eqnarray}
{\cal L}= i \frac{2 g^\prime}{5} \cos\theta_H \overline{T}_- A\slush_H
(\sin \beta P_L + \sin \alpha P_R) t,
\label{lagrangian}
\end{eqnarray}
where,
\begin{eqnarray}
\sin \alpha \simeq \frac{m_t v}{m_{T_-} f}, \ \ \ \ 
\sin \beta \simeq \frac{m_t^2 v}{m_{T_-}^2 f}.
\end{eqnarray}
A top partner $T_-$ decays  dominantly into a top with $h_t=+1/2$
if $m_{T_{-}} \gg m_{A_H} + m_t$ and $\sin\beta \ll \sin\alpha$,
where $h_t$ is defined as the top helicity.
It is the case in our model point, 
since $\sin \beta\simeq m_t/m_{T_-} \sin\alpha \simeq 0.22 \sin\alpha$.
The amplitude is calculated in the Appendix \ref{app_polarization}, 
and we find ${\cal P}\equiv [N(h_t=1/2)-N(h_t=-1/2)]/[N(h_t=1/2)+N(h_t=-1/2)]\sim 0.85$.

To simulate the top polarization effect we need to follow the decay cascade 
till the partons arising from top decay.
Instead of generating
 $pp \to  T_- \overline{T}_- \to b\bar{b} W (\to q\bar{q}) W(\to q\bar{q}) A_H A_H $
 using {\tt COMPHEP}, we generate stop pair production
 $pp \to \ti{t} \ti{t}^\ast$ followed by $\ti{t} \to t \none$ at a MSSM model point  
 using {\tt HERWIG}\footnote{Note that the spin of the intermediate
 particle are different between these two processes.  
Especially, we
expect  spin correlation between $T_- \overline{T}_-$ and their decay products,
which does not exist for $\ti{t}_1 \ti{t}_1^\ast$.
The correlation in principle appear in momentum distribution of $t$ and $\bar{t}$.  
However, 
the effect is rather small because $T_-$ is non-relativistic and also  
the system does not
have enough kinematical constraints.}.
 We take the MSSM parameter  that  the other sparticles are heavy  
 and the
 decay vertex of $\ti{t}_1$ is approximately proportional to $\ti{t}^\ast \none t_R$,
 so that  {\tt HERWIG} generates
 stop pair efficiently and they decay into approximately completely 
 polarized top quark (${\cal P}=0.996$ for our model point).

{\tt HERWIG} has an option to switch off polarization effects.
For the MSSM point, we do not find any distinguishable difference 
between the $m_{T2}$ distributions with/without polarization effects.
Therefore the results shown in Sec.\ref{lht} may be valid 
even for the LHT because the spin correlation effects are not large.

Decay distributions of the top quark contains information on the interaction 
vertex~\cite{CP}. 
The amplitudes for $ h_t=\pm 1/2$ are expressed 
as follows,
\begin{eqnarray}
{\cal M}\sim \sqrt{2m_tE_b}\times 
\begin{cases}
\frac{m_t}{m_W}\cos\frac{\theta}{2} e^{i\phi} &(h_t,\lambda_W, h_b)=(+,0,-),    \\
-\rr\sin\frac{\theta}{2}e^{2i\phi}    &(+,-,-), \\
\frac{m_t}{m_W}\sin\frac{\theta}{2}    & (-,0,-),\\
\rr\cos\frac{\theta}{2}e^{i\phi}    & (-,-,-).
\end{cases}
\label{amp_top}
\end{eqnarray}
The amplitudes for the other helicity combinations vanish.
Here, $E_b=(m_t^2 - m_W^2)/{2 m_t}$ is
the energy of the bottom quark in the rest frame of the decaying top.
And $\theta$ and $\phi$ are 
the polar and azimuthal angles
of the momentum of $W$ boson defined in the rest frame of the decaying top.
The $\theta$ is measured from the top momentum direction
and the $\lambda_{W}$ is a helicity of $W$.
\begin{figure}[htbp]
\rotatebox{-90}{
\includegraphics[scale=0.32]{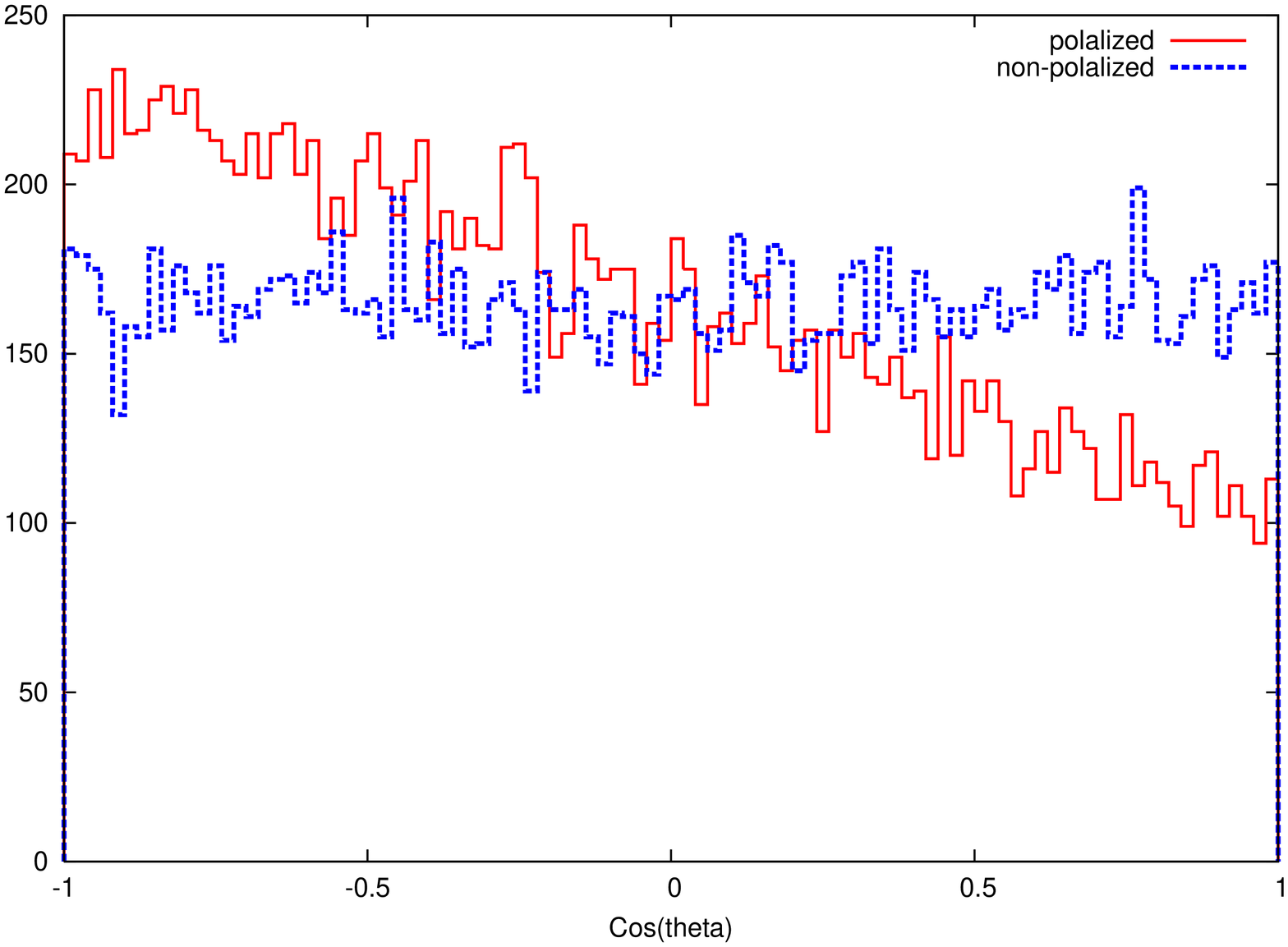}
}
\hfill
\rotatebox{-90}{
\includegraphics[scale=0.32]{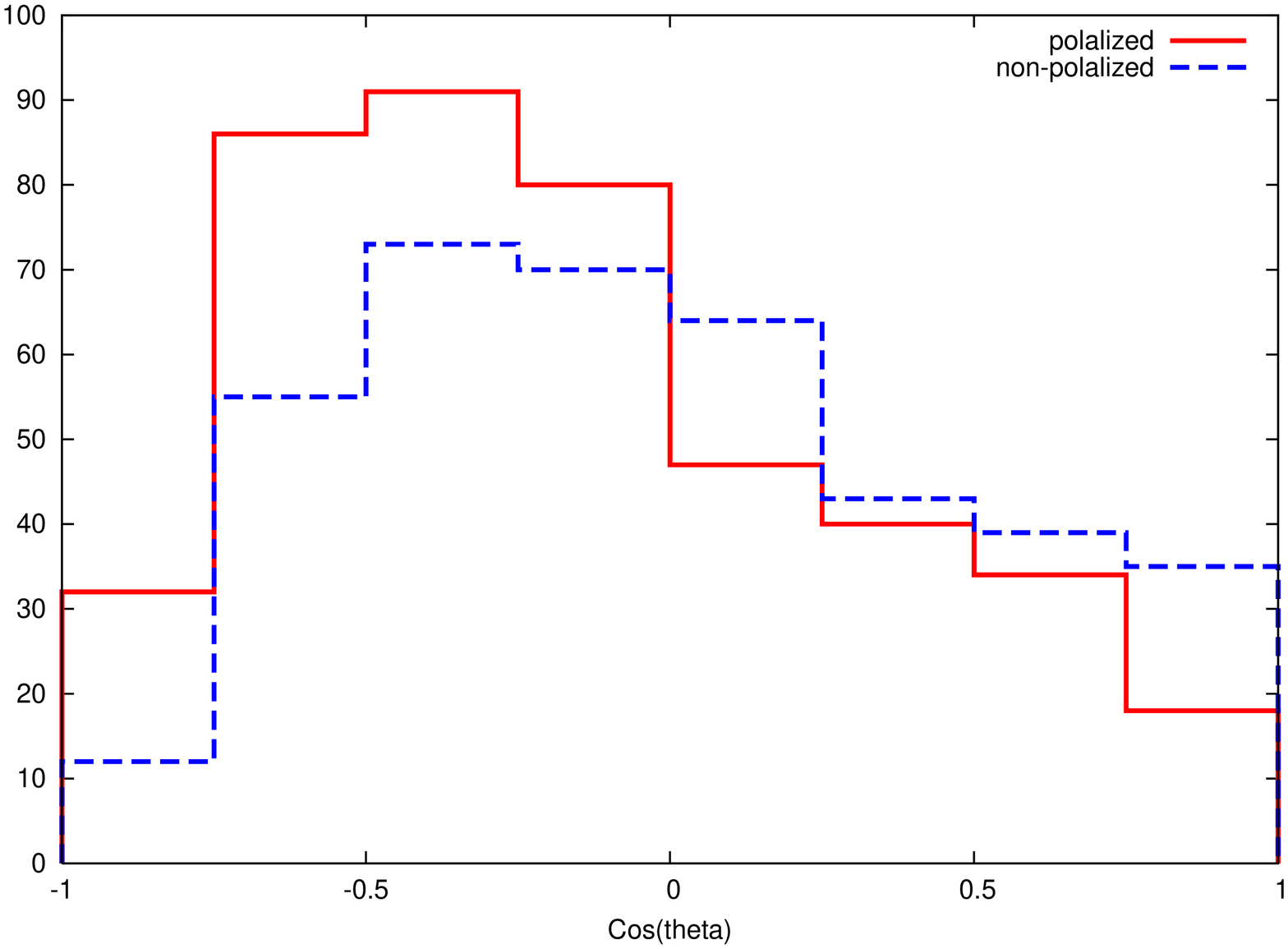}
}
\caption{left) The polar angle ($\cos \theta_{tb}\equiv \cos(\pi - \theta)$) distributions of bottom quarks from polarized (solid) 
and non-polarized (dashed) top decays.
right)
The polar angle ($\cos \theta^{\rm jet}_{tb}$) distributions of the $b$-tagged jets.
}
\label{distribution}
\end{figure}

In Fig.\ref{distribution} (left), we show the parton level top decay distribution 
 as the function of $\cos \theta_{tb}$ for the polarized and non-polarized cases,
where $\theta_{tb}$ is the angle between the $b$-quark momentum and the $t$-quark 
boost direction at the rest flame of the $t$-quark.   
We find  evident difference between the two cases.
In the region around $\cos\theta_{tb} \sim 1$, the emission of $b$
quark is suppressed 
because the amplitude proportional to $\cos (\theta/2) $ is suppressed for $h_t=+1/2$.

To see this at jet level, we study distributions of jets 
that are consistent to top decay products.
Some parton configuration is difficult to resolve at jet level.
We analyse only the hemispheres with
$150$~GeV$< m_{P_H}<190$~GeV.
We require that two of the three jets  are consistent with  those  
coming from $W$, that is $|m_{jj} - m_W|<20\,$GeV, and the other jet is $b$-tagged.
Here, we regard a jet as a $b$-jet if the direction of the momentum is 
in a $\Delta R=0.2$ cone centered at a bottom parton momentum with $p_T>20$~GeV. 
The analysis of the hemisphere which consists of 2 jets
is given in Appendix \ref{app_configuration}.

In Fig.\ref{distribution} (right) we show the distribution of the
angle between the b-jet momentum and the reconstructed top momentum  
$\theta^{\rm jet}_{tb}$.
For the plot, we selected only the events with $m_{T2} > 500\,$GeV.
Under the cut, 
Standard Model backgrounds are negligible as seen in Sec.\ref{lht}.
We use the measured hemisphere momentum to go back to the rest frame of the jet system.
There is a distinguishable difference between polarized and non-polarized distributions.
The ratio $n(\cos\theta^{\rm jet}_{tb} < 0)/ n(\cos\theta^{\rm jet}_{tb} > 0) = 2.08$
for the polarized case, while it is $1.16$ for the unpolarized one.

A polarized top quark decays into a polarized $W$ $(\lambda_W=0,-)$.
Decay distribution of polarized $W$ ($W \to 2j$) can be calculated and the amplitudes are written as follows,
\begin{eqnarray}
{\cal M}_{-}&\propto&\frac{1 - \cos\theta^\ast}{2} e^{-i\phi^\ast}
\cr
{\cal M}_{0}&\propto&\frac{\sin\theta^\ast}{\rr}
\cr
{\cal M}_{+}&\propto&\frac{1 + \cos\theta^\ast}{2} e^{i\phi^\ast}
\end{eqnarray}
Here, $\theta^\ast$ and $\phi^\ast$ are the polar and the azimuth angles of the momentum of one of the jets 
from a $W$ decay to the $W$ momentum direction.
These momenta are defined at the rest frame of the $W$.
A longitudinally polarized $W$ ($\lambda_W=0$) tends to decay transversely.
On the other hand a transversely polarized $W$ ($\lambda_W=\pm$) tends to decay 
along a direction of the $W$ momentum.

\begin{figure}[htbp]
\rotatebox{-90}{
\includegraphics[scale=0.32]{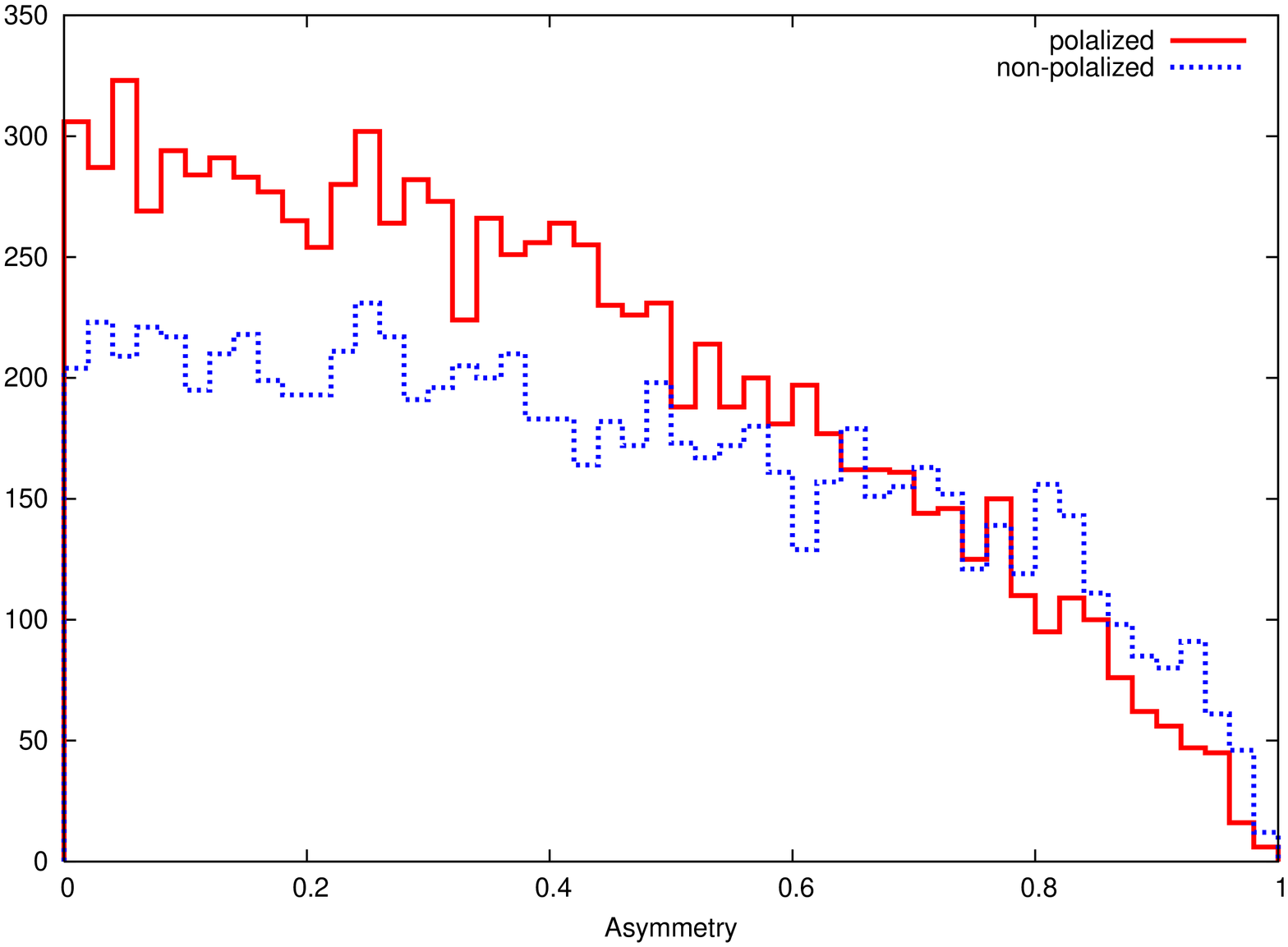}
}
\hfill
\rotatebox{-90}{
\includegraphics[scale=0.32]{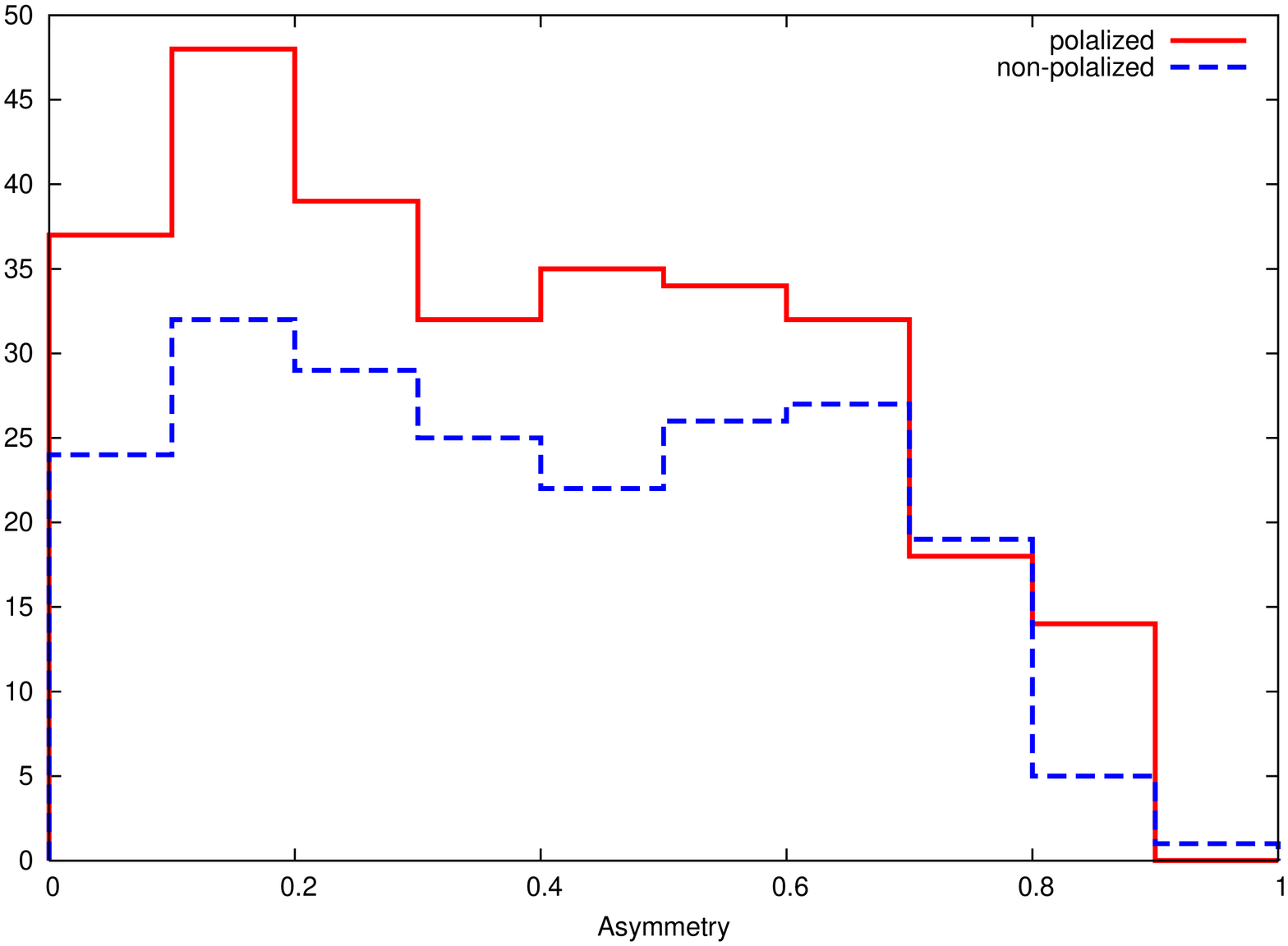}
}
\caption{The distributions of the  $p_T$ asymmetry ${\cal A}$
of the jet pairs from $W$ decays for polarized and non-polarized cases.
Parton level (left) and jet level (right) distributions are shown.
The difference of the distributions can be seen even at jet level.}
\label{asymmetry}
\end{figure}

These differences may appear in the jet $p_T$ asymmetry ${\cal A}$, 
which defined as follows,
\begin{eqnarray}
{\cal A} = \frac{|p_{T1} -p_{T2}|}{p_{T1}+p_{T2}}.
\end{eqnarray}
Jets from a $W(\lambda_W=0)$ decay tend  to have  ${\cal A}\sim 0$ 
while those from a $W(\lambda=\pm 1)$ give larger ${\cal A }$. 
This can be seen in Fig.\ref{asymmetry}~(left). In this 
plot the ${\cal A}_{\rm part}$ distributions are shown 
for the events 
with $\cos\theta^{\rm jets}_{tb}<0$. 
The ratio $N(W(\lambda_W=0))/N(W(\lambda_W=\pm 1))$ 
with $\cos \theta_{tb} <0$ is larger for polarized tops.
Therefore,
${\cal A}$ for polarized top is distributed more around 0  than for non-polarized top.
Fig. \ref{asymmetry}~(right) shows
 ${\cal A}$ distributions at jet level for $\cos \theta^{\rm jet }_{tb}<0$.
Unfortunately, It is difficult to see the differences only by the shape of the distribution
due to the limited statistics.

\section{CONCLUSION}
\label{conclusion}
In this paper we have studied reconstruction of  top quarks 
arising   from $T_- \overline{T}_-$ 
productions and its  subsequent decay into a top and a stable  gauge partner $A_H$ 
in the LHT.   
We demonstrate the reconstruction of the top quarks through finding 
collinear jets whose invariant mass is consistent with $m_t$
using hemisphere analysis.  Main SM background processes are
 $t\bar{t}+$jets, $Z+$jets and $W+$jets productions,
 which can be reduced by imposing the cut on 
 hemisphere momenta and $m_{T2}$ variable. 
 
We also 
investigate the dependence
on jet reconstructing algorithms.
The cone algorithm used in the previous study \cite{Matsumoto:2006ws}
is not optimal for the process.
A top from $T_-$ is boosted, while
the algorithm is designed so that the highest $p_T$ jets 
take all activities near the jet, mis-estimating  the 
energy and the direction of the jets. 
An infrared safe version of the cone algorithms (SISCone) 
also has some disadvantage for our case, because they tend 
to merge overlapping jets.
We also study distributions with modern 
clustering algorithms (kt and Cambridge), which in general give better results than the 
cone algorithms.

We also study effect of underlying events, 
and find the known tendency 
that the kt algorithm overestimates jet energies caused 
by collecting far and soft activities. 
Whereas the reconstruction efficiency in the Cambridge algorithm is not affected if $R\sim 0.4$. 

We also discuss top polarization effects.   A top quark arising 
 from a top partner decay  is naturally polarized. 
This can be studied through looking into a distribution of the  $b$-jet from $t$ decay 
especially the angle to a reconstructed top momentum in the rest frame of the top. 
We find that 
difference of the distributions between 
polarized  and non-polarized top is still at  detectable
level  with  $b$-tagging for reasonable integrated luminosity 
($\int{\cal L}=$50 fb$^{-1}$ for $m_{T_-}=800$GeV and $m_{A_H}=150$GeV).
These analyses are demonstrated using the Cambridge algorithm, 
which shows good $b$-jet and $b$-parton matching.  

In many new physics  scenarios, boosted gauge bosons and 
top quarks are produced at a significant rate. 
Our study shows that choosing a right jet reconstruction 
algorithm 
or studying the dependence on them
is important to reveal the physics behind the signal.  

\section*{Acknowledgement}
This work is supported in part by the Grant-in-Aid for Science Research, 
Ministry of Education, Culture, Sports, Science and Technology, 
Japan (No.16081207, 18340060 for M.M.N.).
This work is also supported by World Premier International Research Center Initiative (WPI Program), 
MEXT, Japan.

\newpage
\appendix
\section{Appendix}
\subsection{ Jet smearing }
\label{smearing}
In this paper, we show the distributions without jet energy smearing. 
In {\tt AcerDET}, there is an option to smear jet and missing transverse energies.
The smearing is introduced for each jet energy in the Snowmass cone algorithm 
and for a sum of the total transverse momentum (the missing transverse momentum)
rather than for each calorimeter cell.
We do not try to include smearing effects for the other three 
jet reconstruction algorithms (kt, Cambridge, SISCone) in this paper,
because cells that jets consist of depend on the 
jet reconstruction algorithms, therefore comparison of 
smearing effects under the same condition is not easy. 

To obtain a rough idea on the signal and background distributions 
with smearing, we show $m_{T2}$ distributions with/without smearing 
and jet energy calibration in the Snowmass cone algorithm. 
The distribution near the end point is not significantly changed.
The effect is rather small because we take the smearing
based on ATLAS detector performance 50$\times \sqrt{E}$\%.
which is less than 10 \%
 for jet with $p_T>$ 30 GeV.

\begin{figure}[h!]
\includegraphics[scale=0.4]{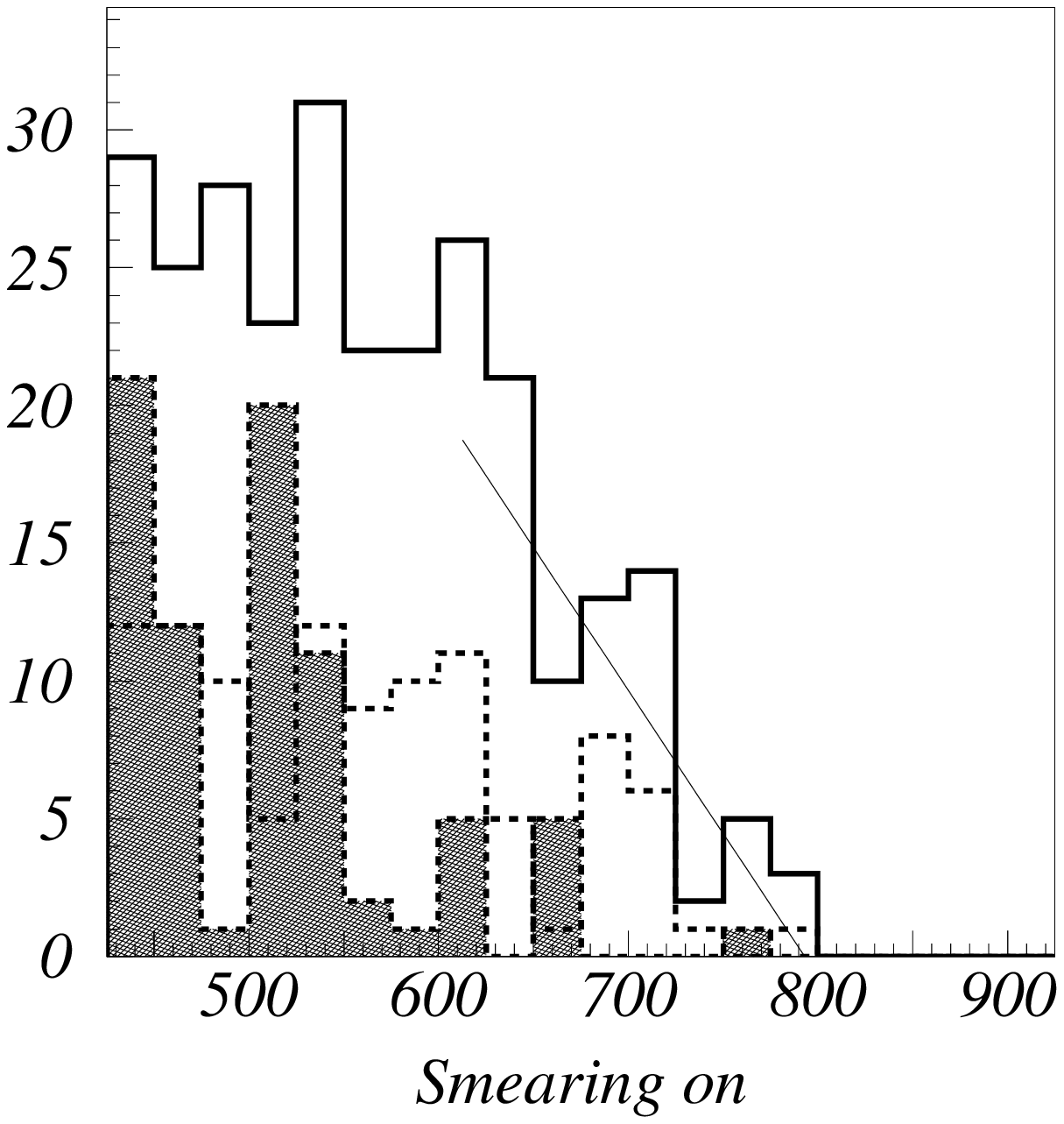}
\includegraphics[scale=0.4]{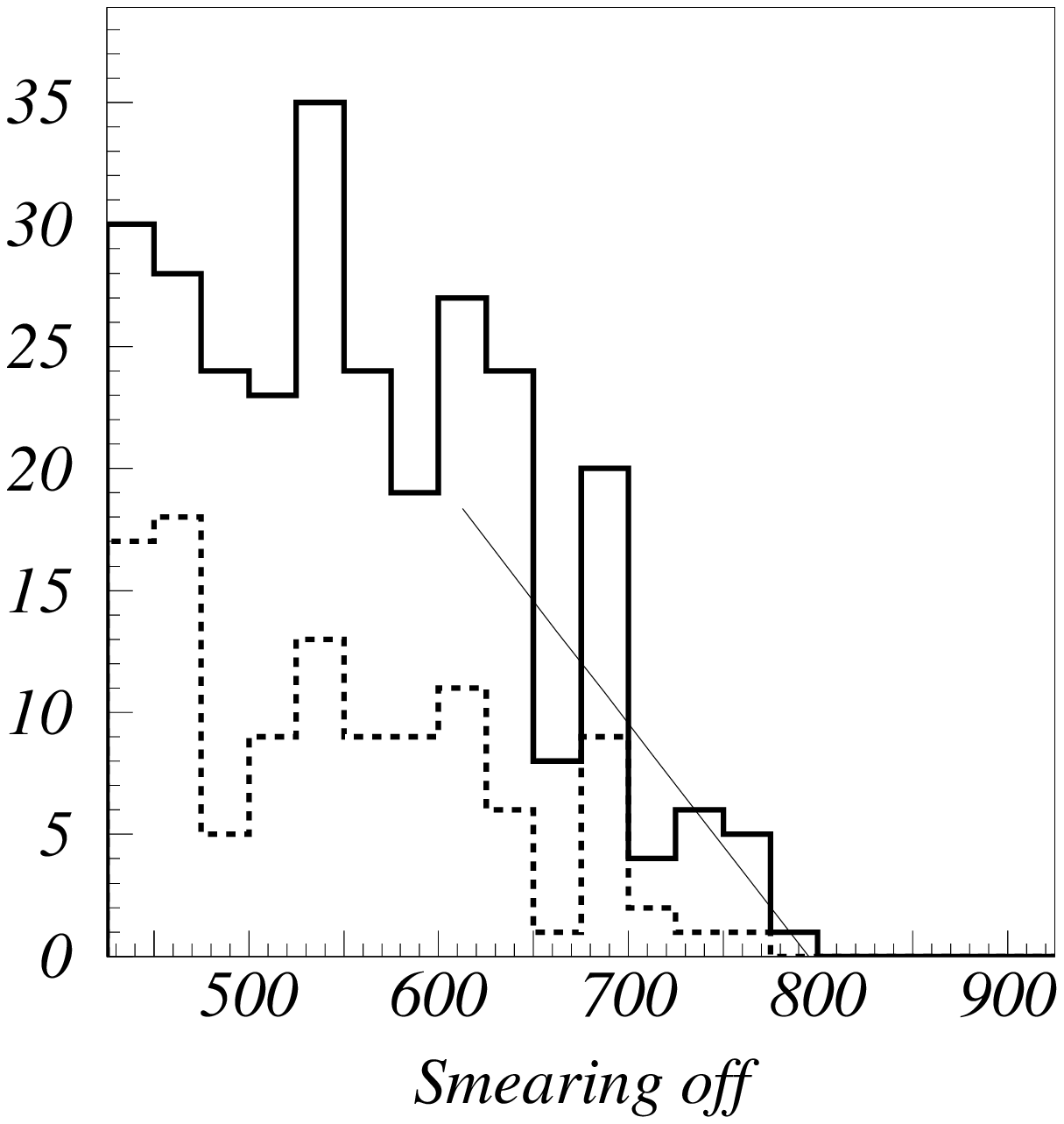}
\caption{Distributions of $m_{T2}$ for the Snowmass cone algorithm without/with smearing (left/right figure respectively).
The dashed line shows for the events with $150{\rm ~GeV} < m_{P_{Hi}}< 190{\rm ~GeV}$ for both $H_i$.
The solid line shows for the events with $50{\rm ~GeV} < m_{P_{Hi}} < 190{\rm ~GeV}$.
The endpoints are 785.9$\pm$6.6/810.2$\pm$25~GeV for nominal
$m_{A_H}=151.8$~GeV .
($m_{T_-}=800.2$~GeV).
Calibrated jets are used for both figures.
Dark histogram is the background distribution in the case that jet energy smearing is on.
We do not show the background distribution without jet energy smearing in the right figure.
}
\label{mt2_snowmass}
\end{figure}

\subsection{Top polarization}
\label{app_polarization}
In this paper we took mass parameters 
$m_{T-}=800.2$~GeV, $m_{A_H}=151.8$~GeV, and $m_t=175.$~GeV.
The Lagrangian relevant to our study is 
\begin{eqnarray}
{\cal L}= i \frac{2 g^\prime}{5} \cos\theta_H \overline{T}_- A\slush_H
(\sin \beta P_L + \sin \alpha P_R) t.
\end{eqnarray}
Here, $\alpha$ and $\beta$ is approximately expressed in terms of
\begin{eqnarray}
\sin \alpha \simeq \frac{m_t v}{m_{T_-} f}, \ \ \ 
\sin \beta \simeq \frac{m_t^2 v}{m_{T_-}^2 f},
\end{eqnarray}
therefore $\sin \beta\simeq 0.22 \sin\alpha$ at our model point.

The amplitude of a top partner decay into a top with helicities $h_t,h_T,\lambda_A$ can be calculated as follows, 
\begin{eqnarray}
{\cal M}_{h_t,h_T,\lambda_A} &\sim& -i<t A_H | \bar{t} A\slush _H (s_\beta P_L + s_\alpha P_R) T_- |T_->
\cr&=& -i \epsilon^{\ast}_{h_A\mu} (\mathbf{p}_{A_H};m_{A_H})
 \bar{u}_{h_t}(\mathbf{p}_t;m_t)
 \gamma^\mu (s_\beta P_L + s_\alpha P_R)
 u_{h_T}(\mathbf{p}_{T_-};m_{T})
\cr
&=&
-ie^{i(h_T-h_t+\lambda_A)\phi} \sqrt{2m_{T_-}} \left[
- \gamma_A \beta_A \delta_{\lambda_A 0}
 d^\h_{h_t,h_T}(\theta)
(\frac{s_\alpha + s_\beta }{2}
A_+^t + 
2 h_t\frac{s_\alpha - s_\beta }{2} 
 A_-^t) 
\right.
\cr
&&\left.
- \sqrt{2}^{|\lambda_A|} 
\gamma_A^{1-|\lambda_A|} d^\h_{\lambda_A+h_T,h_t}(\theta)
(\frac{s_\alpha + s_\beta }{2}
 A_-^t
+ 2 h_t \frac{s_\alpha - s_\beta }{2} 
 A_+^t )
\right],
\end{eqnarray}
where at the rest frame of $T_-$,
\begin{eqnarray}
\mathbf{p}_t &=&(p_t\sin\theta \cos\phi,p_t\sin\theta\sin\phi,p_t \cos\theta)
= - \mathbf{p}_{A_H},
\cr
E_t&=&\frac{m_{T_-}}{2}+\Delta, \ \ \ \ 
E_{A}=\frac{m_{T_-}}{2}-\Delta,\\
\Delta&=&\frac{m_t^2-m_{A_H}^2}{2 m_{T_-}},\\
\cr
p_t&=&p_{A_H}=\frac{m_{T_-}}{2}
 \sqrt{1-2\frac{m_t^2+m_{A_H}^2}{m_{T_-}^2} + \frac{(m_t^2-m_{A_H}^2)^2}{m_{T_-}^4}},
\end{eqnarray}
and, 
\begin{eqnarray}
A_\pm^t&=&\sqrt{E_t \pm m_t},\\
\gamma_{A}&=& E_{A_H}/m_{A_H},\ \ \ \ 
\beta_{A}= p_{A_H}/E_{A_H}.
\end{eqnarray}
The $d^\h_{h_1,h_2}(\theta)$ is the Wigner's function,
\begin{eqnarray}
d^\h_{h_1,h_2}(\theta)=
\bordermatrix{
_{h_1} \setminus^{h_2}&\h &-\h \cr
\h& \cos \frac{\theta}{2}& -\sin\frac{\theta}{2}\cr
-\h&\sin\frac{\theta}{2} &  \cos \frac{\theta}{2} }.
\label{dfun}
\end{eqnarray}
The square of the amplitudes are therefore expressed for each helicity eigenstate 
as follows;
\begin{eqnarray}
|{\cal M}|^2\sim 2 m_{T_-}
\begin{cases}
2\cos^2 \frac{\theta}{2}
(\frac{s_\alpha + s_\beta }{2} A_-^t +  \frac{s_\alpha - s_\beta }{2}  A_+^t )^2,
&(+,-,+),\cr
2\sin^2 \frac{\theta}{2}
(\frac{s_\alpha + s_\beta }{2} A_-^t +  \frac{s_\alpha - s_\beta }{2}  A_+^t )^2,
&(+,+,-),\cr
\gamma_{A}^2 \cos^2{\frac{\theta}{2}}
[\beta_{A} (\frac{s_\alpha + s_\beta }{2}A_+^t +\frac{s_\alpha - s_\beta }{2}  A_-^t) 
+ (\frac{s_\alpha + s_\beta }{2} A_-^t +  \frac{s_\alpha - s_\beta }{2}  A_+^t )]^2,
&(+,+,0),\cr
\gamma_{A}^2 \sin^2\frac{\theta}{2}
[\beta_{A} (\frac{s_\alpha + s_\beta }{2}A_+^t +\frac{s_\alpha - s_\beta }{2}  A_-^t) 
- (\frac{s_\alpha + s_\beta }{2} A_-^t +  \frac{s_\alpha - s_\beta }{2}  A_+^t )]^2,
&(+,-,0),\cr
2\sin^2 \frac{\theta}{2}
(\frac{s_\alpha + s_\beta }{2} A_-^t - \frac{s_\alpha - s_\beta }{2}  A_+^t )^2,
&(-,-,+),\cr
2\cos^2 \frac{\theta}{2} 
(\frac{s_\alpha + s_\beta }{2} A_-^t -  \frac{s_\alpha - s_\beta }{2}  A_+^t )^2,
&(-,+,-),\cr
\gamma_{A}^2 \sin^2 \frac{\theta}{2} 
[\beta_{A} (\frac{s_\alpha + s_\beta }{2}A_+^t -\frac{s_\alpha - s_\beta }{2}  A_-^t) 
- (\frac{s_\alpha + s_\beta }{2} A_-^t -  \frac{s_\alpha - s_\beta }{2}  A_+^t )]^2,
&(-,+,0),\cr
\gamma_{A}^2 \cos^2 \frac{\theta}{2}
[\beta_{A} (\frac{s_\alpha + s_\beta }{2}A_+^t -\frac{s_\alpha - s_\beta }{2}  A_-^t) 
+ (\frac{s_\alpha + s_\beta }{2} A_-^t -  \frac{s_\alpha - s_\beta }{2}  A_+^t )]^2,
&(-,-,0),\cr
0 &({\rm the\ others}). \cr
\end{cases}
\end{eqnarray}

Polarization of top quarks from top partner decays is given as follows,

\begin{eqnarray}
{\cal P} &=& 
\frac{\Gamma(T_- \to t_{(+)} A_H) - \Gamma(T_- \to t_{(-)} A_H)}
{\Gamma(T_- \to t_{(+)} A_H) + \Gamma(T_- \to t_{(-)} A_H)}
\cr
\cr
 &=&
\frac{\overline{\sum}_{h_T,\lambda_A}|{\cal M}_{h_t=+\h,h_T,\lambda_A}|^2-
\overline{\sum}_{h_T,\lambda_A}|{\cal M}_{h_t=-\h,h_T,\lambda_A}|^2}
{\sum_{h_t} \overline{\sum}_{h_T,\lambda_A}|{\cal M}_{h_t,h_T,\lambda_A}|^2}
\cr
\cr
 &=&
\frac{ (s_\alpha^2 -s_\beta^2)|p_t|(2+ \gamma_{A}^2+\gamma_{A}^2\beta_{A}^2 )}
{(s_\alpha^2+s_\beta^2) (2+\gamma_{A}^2 + \gamma_{A}^2\beta_{A}^2)E_t
- 2s_\alpha s_\beta (2+\gamma_{A}^2 - \gamma_{A}^2\beta_{A}^2)m_t }. 
\end{eqnarray}

We obtain ${\cal P}\sim 0.85$ for our model parameter.

\subsection{Jets Configurations in Hemispheres}
\label{app_configuration}

\begin{figure}[h!tbp]
\rotatebox{-90}{
\includegraphics[scale=0.32]{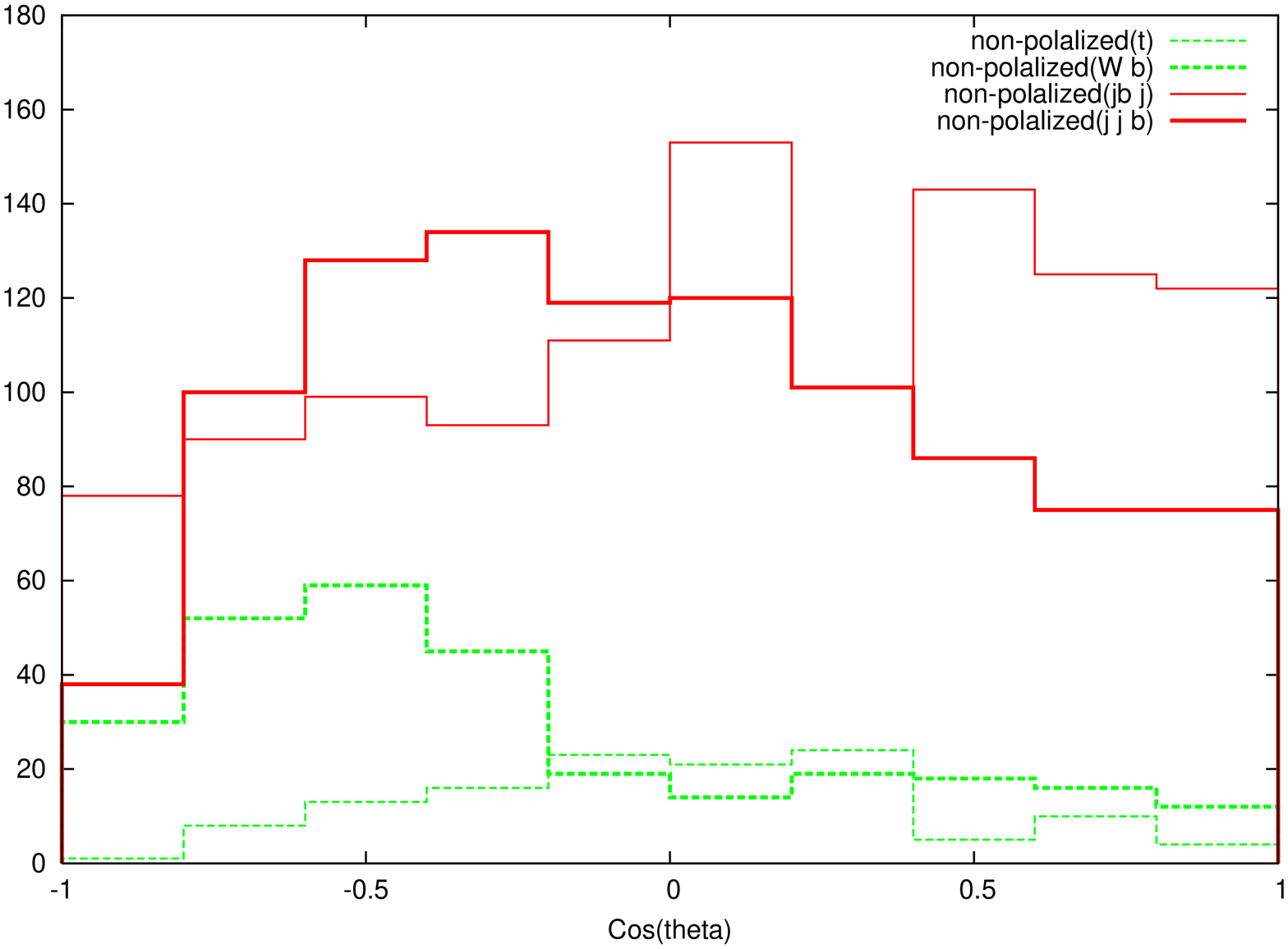}
}
\hfill
\rotatebox{-90}{
\includegraphics[scale=0.32]{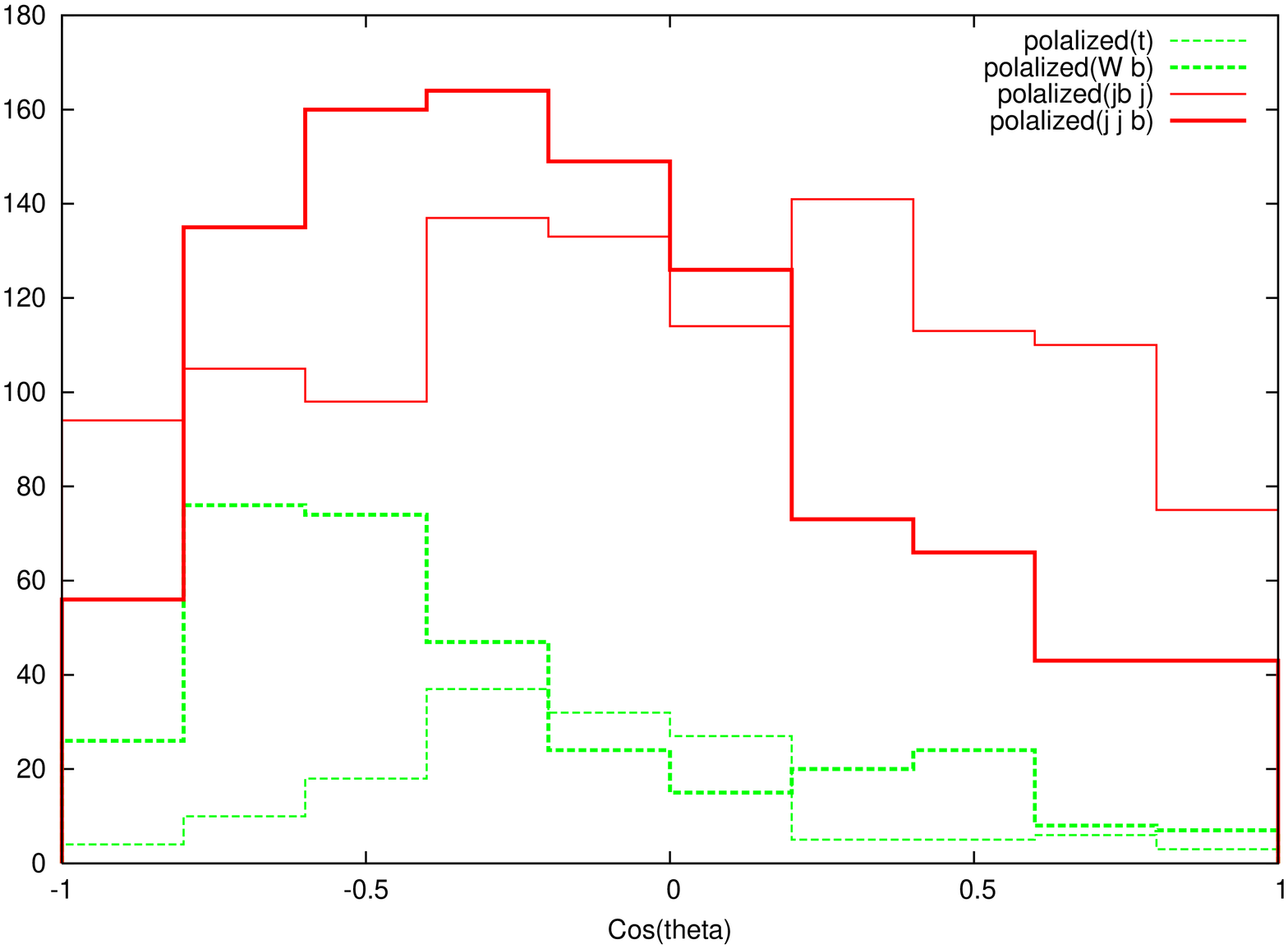}
}
\caption{The distributions of the $b$-partons in hemispheres
that the total hemisphere mass is consistent with 
$m_t$  as the function of  $\theta_{tb}$. 
The right(left)  figure is  for (non-)polarized tops. 
Distributions are shown for the four different groups. 
I) one jets in the hemisphere (dashed thin line).
II) two jets in the hemisphere  and one jet has mass consistent with $W$ (dashed thick line).  
III) two jets in the hemisphere  but none of jets has mass consistent with $W$ (solid thin line). 
IV)  three jets in the hemisphere 
 and mass of a pair of two jets is consistent with $W$ (solid thick line).}
\label{typehemi}
\end{figure}

For hemispheres consistent with a top mass,
some parton configurations are difficult to resolve at jet level.
We categorize hemispheres  with 
$150$~GeV$< m_{P_H}<190$~GeV
into the following four groups:
\begin{description}
\item{I.}  Only one jet in a hemisphere.
\item{II.}  Only two jets in a hemisphere. One of the jets has 
mass consistent with $W$($m_{W}-20$~GeV $<m_j<m_W+20$~GeV) and the other is $b$-tagged.  
($W$ decay products  are merged into a jet and $b$-jet is isolated)
\item{III.}  Only two jets in a hemisphere. None of the jets has
mass consistent with $W$ mass and at least one of jets is $b$-tagged.
(One of the partons  from $W$ decay and a $b$-parton  are merged into a jet, 
and the other from $W$ is isolated. )
\item{IV.}  Only three jets in a hemisphere.  
Invariant mass of two jets is consistent with $W$ and the other jet is $b$-tagged.
\end{description}
Here, we regard a jet as a $b$-jet if the direction of the momentum is in a $\Delta R=0.2$ cone centered at a
bottom parton momentum with $p_T>20$~GeV. 

In Fig.~\ref{typehemi},  we plot $\theta_{tb}$ distribution of jets 
for each group,
where $\theta_{tb}$ is the angle between a $b$-parton and a $t$-parton momenta
at the rest frame of the $t$-parton.
The left figure is for non-polarized top quarks and the right figure 
is for polarized top quarks.

For type IV events,
the number of the events where a $b$-parton goes in the forward direction is strongly suppressed
for polarized events compared with non-polarized events, 
while the events for $\cos\theta_{tb}<0$ is significantly enhanced.
It is consistent with parton level distributions.
Note that the events near $\cos\theta_{tb}\sim -1$ cannot be accepted as type IV for both polarized and non-polarized cases
because $p_T$ of the $b$-parton is too small, 
while the region is the most sensitive to polarization effects.
The other distributions do not show clear dependence on the polarization.

\end{document}